\documentclass[a4paper,11pt]{article}
\usepackage{amsmath, amssymb}
\usepackage[margin=1in]{geometry}
\usepackage{color, bm,graphicx}
\usepackage{cite}

\newcommand{\email}[1]{\thanks{email: #1}}

\newcommand{\be}{\begin{equation}}
\newcommand{\ee}{\end{equation}}

\begin{document}

\title{Renormalization, Decoupling and the Hierarchy Problem}
\author{Kang-Sin Choi\email{kangsin@ewha.ac.kr}
 \\ \it \normalsize 
Scranton Honors Program, Ewha Womans University, Seoul 03760, Korea \\
\it \normalsize  
Institute of Mathematical Sciences, Ewha Womans University, Seoul 03760, Korea}
\date{}

\maketitle

\begin{abstract}
The hierarchy problem is associated with renormalization and decoupling.
The smallness of the scalar mass against loop corrections follows from the decoupling of heavy fields, once the observable is correctly identified as the renormalized mass depending on the external momentum rather than a constant parameter. We reconsider the properties of the renormalized loop corrections, which are finite, independent of regularization and admit a well-defined perturbative expansion, so that a quadratic dependence on a regulator cutoff carries no physical
information. 
The substantive question is the dependence on a physical heavy mass $M$, and we show, by explicit calculation up to two loops and by power counting to all orders, that the renormalized correction to the scalar mass-squared from a heavy field is suppressed as $(p^2-m_h^2)^2/M^2$, where $p$ is the external momentum and $m_h$ the scalar pole mass, provided the couplings
are held fixed as $M \to \infty$.
The surviving momentum dependence is in principle measurable. This is in accordance with the Appelquist--Carazzone decoupling theorem, whose statement holds up to the renormalization of the light masses and which we make explicit for the scalar mass. We do not address the origin of the small tree-level mass.
\end{abstract}

\newpage

\section{Introduction and summary}

In quantum field theory, a physical quantity is corrected by loop amplitudes.  
To the bare mass-squared parameter $m_B^2$ of a scalar field, e.g., the Higgs, loop corrections are added as
\be \label{massintro}
 m_B^2 + \sum_{\{l_i\}} \Sigma_{l_1,l_2,\dots,l_n}(p^2), 
\ee
where 
\be \label{prop}
 \Sigma_{l_1,l_2,\dots,l_n}(p^2)  \propto g_1^{l_1} g_2^{l_2}\dots g_n^{l_n}, \quad l_i = 0,1,2,\dots
 \ee 
are the so-called self-energies with the external momentum $p$, characterized by their one-particle-irreducible (1PI) Feynman diagrams with two external legs of the scalar. The subscripts denote the dependence on the couplings $g_1, g_2, \dots, g_n$, including gauge couplings, Yukawa couplings and so on.

Scalar field theory suffers from the hierarchy problem \cite{Choi:2023cqs, Choi:2023mma, Georgi:1974yw, Gildener:1976ai, Gildener:1979dd,Weinberg:1978ym, Veltman:1980mj,Natale:1982mt,Susskind:1982mw,Wetterich:1983bi} (see also \cite{Farina:2013mla,Feng:2013pwa, Hamada:2012bp,Wells:2013tta, Hebecker:2020aqr, Mooij:2021ojy, Aoki:2012xs, Jegerlehner:2013cta, Jegerlehner:2021ckr, Wetterich:2011aa}). It has been known that each correction $\Sigma_{l_1,l_2,\dots,l_n}(p^2)$ diverges quadratically, scaling as $\Lambda^2$, where $\Lambda$ is the momentum cutoff of the loop integral. 
However, $\Lambda$ is not a parameter of the theory. 
We show that, since the scalar mass is renormalizable, the dependence on $\Lambda$ is removed by renormalization and carries no physical information.

The correction also depends on physical, dimensionful parameters, such as the mass $M$ of a field in ultraviolet (UV) physics. For example, $M$ can be the mass of the heaviest field running in the loop. In general, there
may be multiple fields and $M$ may be defined as the average or representative mass scale. By dimensional analysis, each correction $\Sigma_{l_1,l_2,\dots,l_n}(p^2)$ is proportional to $M^2$ and is dominated by the heaviest field, which suggests that the scalar mass is sensitive to unknown UV physics and not well-defined at low energies. 
No argument about the regulator removes this, and this is where the problem is substantive. We therefore reserve the term {\em technical hierarchy problem} for this question, whether a small pole mass, once imposed, is ruined by loop corrections of order $M^2$ from heavy fields. 
This is technical naturalness in the sense of 't Hooft \cite{tHooft:1979rat}, namely the radiative stability of a hierarchy that has already been arranged, and it is the counterpart of the ``big'' hierarchy problem, which asks where the small tree-level mass comes from in the first place. The latter we {\em do not} attempt to address, as it does not give rise to
a consistency problem.

This paper points out that the technical part of the hierarchy problem,
arising from the loops, is related to renormalization and decoupling. What
we show is that, once the pole mass is made small, it {\em remains small
against loop corrections}, because the heavy fields decouple.

It is crucial to identify the observable physical parameters as the renormalized ones. Although the bare parameters define the theory, the bare mass $m_B^2$ can never be observed because interactions modify it \cite{Choi:2023cqs, Choi:2023mma}; hence, only the whole combination of the mass (\ref{massintro}) can be observed.
Thus, it is {\em not necessary} to make the {\em unrenormalized} loop corrections $\Sigma_{l_1,l_2,\dots,l_n}(p^2)$ in (\ref{massintro}) {\em small}, as has been attempted in the usual formulation. The common premise that $m_B^2$ is small is not necessarily true, either. Note the renormalization condition
\be \label{baremassagain}
 m_h^2 = m_B^2 + \sum_{\{l_i\}}  \Sigma_{l_1,l_2,\dots,l_n}(m_h^2),
\ee
where $m_h^2$ is the pole mass defined at the renormalization point $p^2=m_h^2$. From this, we see that for small $m_h^2$, the other two in (\ref{baremassagain}), the bare mass and the sum of the self-energies, should be comparable. 

At first sight, the smallness of the observed mass (\ref{massintro}) requires fine-tuning between the two huge parameters. However, the cancellation is natural if we rewrite the bare mass in (\ref{massintro}) using the relation (\ref{baremassagain}). We can {\em define the same theory} using the mass $m_h^2$ instead of $m_B^2$ through (\ref{baremassagain}). In effective field theory, the parameters in the Lagrangian quantify our ignorance, and we fit them to experiments; a change of basis neither gains nor loses information. 

One of the obstacles to the hierarchy program has been the identification of the renormalized mass. This also requires field-strength renormalization \cite{Dyson:1949ha, BS55}, which introduces additional terms that we will see below.
Then, the Higgs mass admits a well-defined perturbative expansion
\be \label{totmassub} 
\begin{split}
m^2(p^2) &= m_h^2+ \sum_{\{l_i\}} \Sigma^{\text{ren}}_{l_1,l_2,\dots,l_n}(p^2) \\
 & \equiv m_h^2+ \sum_{\{l_i\}} \left[  \Sigma_{l_1,l_2,\dots,l_n}(p^2) -\Sigma_{l_1,l_2,\dots,l_n}(m_h^2) - (p^2-m_h^2) \frac{d\Sigma_{l_1,l_2,\dots,l_n}}{d(p^2)}(m_h^2) \right ].
\end{split}
\ee
That is, the bare mass is the constant part of $m^2(p^2)$ in (\ref{totmassub}). This combination, which depends on the {\em external momentum,} is the effective mass that we can only observe.

It was shown by Zimmermann \cite{Zimmermann:1969jj} that each term in the second line in (\ref{totmassub}) is finite, even if we extend the momentum integral to infinity. This means the total mass $m^2(p^2)$ is independent of the cutoff $\Lambda$ and, in general, regularization. There is no need for a miraculous cancellation.
The combination (\ref{totmassub}) should be finite because the same is obtained by running between the two scales, from $m$ to $p$, which is purely low-energy behavior \cite{Choi:2023mma}.
If we wrote the Lagrangian using (\ref{baremassagain}), there is no problem of infinity.

Our last concern is whether such loop corrections depend on the UV parameters \cite{Martin:1997ns, Giudice:2013yca, Cohen:2019wxr, Craig:2022eqo}. For instance, one may worry that a one-loop correction by a very heavy field with the mass $M$ can change a coupling. Then, the Appelquist--Carazzone decoupling theorem states that the corresponding amplitude is suppressed in some powers of $p^2/M^2$ and $m_h^2/M^2$ \cite{Appelquist:1974tg}. In other words, the renormalized loop corrections vanish in the limit $M \to \infty$.

We show that this decoupling theorem also holds for the scalar mass $m^2(p^2)$. That is, the effect on a UV field vanishes in the limit of its heavy mass \cite{Choi:2023cqs, Choi:2023mma}
\be
  \quad M \to \infty \quad \Longrightarrow \quad \Sigma^{\text{ren}}_{l_1,l_2,\dots,l_n}(p^2) \to 0,
\ee
if this particular mass correction involves the field of the heavy mass $M$.
This justifies the claim that the scalar mass ought to be insensitive to UV physics, as required for a well-defined effective field theory below a certain scale.

Our argument has two parts, corresponding to the two dependences separated above. The bare mass $m_B^2$ is not observable, so its apparent cancellation against the loop corrections in (\ref{baremassagain}) reflects how we organize the theory rather than fine-tuning between independent physical quantities. We can define the same theory using the pole mass $m_h^2$, from which the bare mass is derived through (\ref{baremassagain}), so the cancellation question does not arise at the level of Lagrangian parameters. The second part concerns the size of the renormalized loop corrections themselves. $\Sigma^{\text{ren}}_{l_1,l_2,\dots,l_n}(p^2)$ from a heavy field of mass $M$ is parametrically small, scaling as $(p^2-m_h^2)^2/M^2$. Heavy physics decouples from the physical mass at the same level it decouples from couplings via the Appelquist--Carazzone theorem. The first statement is about how we organize the calculation; the second is about the actual size of loop corrections to the physical observable. Together, they resolve the technical part of the hierarchy problem without invoking new physics or fine-tuning. The contrast with the $\overline{\rm MS}$ matching procedure, in which threshold matching coefficients separately contain $M^2 \log M^2$ pieces that must be reabsorbed at each scale, is discussed in Sec.~\ref{subsec:intermscale}.

%%%%%%%%%%%%%%%%%%%%%%%%%%%%%%%%%%%%%%%%%
\section{Renormalization using counterterms}

We briefly review the conventional renormalization using counterterms.What renormalization really teaches us is that a physical quantity is scale-dependent, the scale being set by the external momentum $p^2$ at which it is probed. This is the point of view we keep throughout, in contrast to the renormalization group scale, usually denoted by $\mu$, which is not itself an observable.

\subsection{On-shell renormalization}

We are mainly interested in an example of the $\phi^4$-theory, which appears as the Higgs sector in the Standard Model. The Lagrangian density is
\be \label{phi4Lag}
 {\cal L} = \frac12 \partial_\mu \phi \partial^\mu \phi - \frac12 m_B^2 \phi^2 - \frac{6\lambda_B}{24} \phi^4.
\ee
The mass $m_B^2$ and the quartic coupling $\lambda_B$ are bare parameters defining the theory. 

If we calculate the actual scattering amplitude, loop corrections modify the observed parameters. The number of loops corresponds to powers of the Planck constant $\hbar$, reflecting the quantum nature \cite{Nambu:1968rr}. Such corrections typically diverge, which is to be remedied as follows.

We re-normalize the field
\be \label{fieldstr}
 \phi = \sqrt{Z} \phi_r
\ee
to {\em rewrite} the Lagrangian (\ref{phi4Lag})
\be \label{renLag}
\begin{split}
 {\cal L} &= \frac12 \partial_\mu \phi_r \partial^\mu \phi_r - \frac12 m_h^2 \phi_r^2 - \frac{6\lambda}{24} \phi_r^4 
  + \frac12 \delta_Z \partial_\mu \phi_r \partial^\mu \phi_r - \frac12 \delta_{m} \phi_r^2 - \frac{6\delta_\lambda}{24} \phi_r^4,
\end{split}
\ee
with
\be 
 \delta_Z = Z-1, \quad \delta_{m} = m_B^2 Z- m_h^2,\quad \delta_\lambda = \lambda_B Z^2 - \lambda.
\ee
We are free to choose expansion parameters and their reference points.
In the on-shell (OS) or physical scheme (see, e.g., \cite{Sirlin:1980nh, Peskin:1995ev, Weinberg:1995mt}), we regard the parameters $m_h^2$ and $\lambda$ in (\ref{renLag}) as the observed ones through the scattering process at a reference scale, to be discussed below. 
The (Feynman) propagator contains this mass $m_h^2,$
\be \label{FeynProp}
 D(p^2) = \frac{i}{p^2- m_h^2 }.
\ee
The remaining terms of $\delta_Z, \delta_m, \delta_\lambda$ are so-called counterterms.
They are going to absorb the divergences from the loop amplitude. If we instead add counterterms on top of the bare Lagrangian, we redefine the theory.
Due to the symmetry factors and the normalization in (\ref{renLag}), the natural expansion parameters in this paper are $6\lambda$ and $6\delta_\lambda$.

In quantum theory, the mass-squared parameter is always modified by the self-energy $\Sigma(p^2)$, the total 1PI amplitude with two external scalars. There are contributions from the counterterms as well,
\be \label{denom}
 i D^{-1} (p^2)=  p^2 - m_h^2 - \Sigma(p^2) + p^2 \delta_{Z} - \delta_{m}.
\ee
In the OS scheme, we impose a renormalization condition in which the mass used in the propagator remains the same. That is, the correction to the propagator should be 
\be \label{MRenCond}
 \left[ \Sigma(p^2) - p^2 \delta_{Z} + \delta_{m} \right]_{p^2=m_h^2} = 0,
\ee
to all orders of perturbation. For the momentum-dependent interaction $p^2 \delta_Z$, we impose a further condition that the field $\phi_r$ should be approximately free\footnote{Here and in what follows, $d/dp^2$ means differentiation with respect to $p^2$.}  to ${\cal O}(p^2-m_h^2)$
\be \label{ZRenCond}
 \left[ \frac{d \Sigma}{d p^2}(p^2) - \delta_{Z} \right]_{p^2=m_h^2} = 0.
\ee

From dimensional analysis, the self-energy $\Sigma(p^2)$ is a quadratic polynomial in $\sqrt{p^2}$. The constant $\delta_{m}$ is chosen to keep the total mass at the specific value at $p^2=m_h^2$ unchanged. At the same time, it takes away the quadratic divergence in $\Sigma(p^2)$. Our theory has parity symmetry, so there is no term linear in $p$.
The constant $\delta_Z$ absorbs the logarithmic divergence in the coefficient of $p^2$ in $\Sigma(p^2)$.

The self-energy has formal expansion in the coupling $\lambda$, 
\be \label{selfE}
\Sigma(p^2) = \sum_{l=1}^\infty \Sigma_l(p^2),\quad \Sigma_l(p^2) \propto \lambda^l.
\ee 
Also, the counterterm parameters have a formal expansion
\cite{Coleman:2018mew}
\be \label{CTexp}
\begin{split}
 \delta_Z &= \sum_{l=1}^\infty (6\lambda)^l \delta_{Z l}=\sum_{l=1}^\infty \frac{d \Sigma_l}{d p^2}(m_h^2), \\
  \delta_m &= \sum_{l=1}^\infty (6\lambda)^l \delta_{m l}= \sum_{l=1}^\infty \left(- \Sigma_l(m_h^2) + m_h^2 \frac{d \Sigma_l}{d p^2}(m_h^2)\right), \\
   6\delta_\lambda &= \sum_{l=1}^\infty (6\lambda)^{l+1}  \delta_{\lambda  l} %=\sum_{l=1}^\infty  Q_l(p)
   ,
\end{split}
\ee
from the conditions (\ref{MRenCond}) and (\ref{ZRenCond}).

%%%%%%%%%%%%%%%%%%%%%%%
\subsection{Prediction from renormalization}

What is the prediction if all the quantum corrections to the propagator do not modify the pole mass by default? We have fixed the counterterms to order $p^2$ in (\ref{ZRenCond}) that can be rewritten in the Lagrangian. In terms of the renormalized field $\phi_r$, the quadratic Lagrangian is rewritten as
\be \label{altdef}
\begin{split}
 {\cal L}_{\text{quad}} &= \frac12 \left(1+\frac{d \Sigma}{ dp^2}(m_h^2) \right) \partial_\mu \phi_r \partial^\mu \phi_r - \frac12 \left( m_h^2- \Sigma(m_h^2)+m_h^2 \frac{d \Sigma}{ dp^2}(m_h^2)  \right) \phi_r^2 
\end{split}
\ee
or in the momentum space,
\be 
\hat {\cal L}_{\text{quad}} =\frac12 \left(p^2 - m_h^2 +\Sigma(m_h^2) + (p^2-m_h^2)\frac{d\Sigma}{d p^2}(m_h^2)\right)  \phi_r(-p) \phi_r(p).
\ee 
The counterterms not only remove the divergences but also {\em contribute to the physical quantity.}
In the scattering experiment,  due to the superposition of all possible interactions, always the combination
\be \label{gamma2}
\begin{split}
  \Gamma^{(2)}(p^2) &= p^2 - m_h^2 - \Sigma(p^2) + \Sigma(m_h^2) + (p^2 -m_h^2) \frac{d\Sigma}{d p^2} (m_h^2)\\
  & = \langle \phi_r (-p) \phi_r(p) \rangle_{\text{1PI}}
\end{split}
\ee 
and its inverse appears in the $S$-matrix. Taking into account the quantum correction and comparing it with the free Green's function, it is reasonable to read off the mass as \cite{Coleman:2018mew}
\be \label{SlidingMass} 
\begin{split}
 m^2(p^2) &\equiv p^2 - \Gamma^{(2)}(p^2) \\
 &= m_h^2 + \Sigma(p^2) - \Sigma(m_h^2) - (p^2 -m_h^2) \frac{d\Sigma}{d p^2} (m_h^2).
\end{split}
\ee
It follows that {\em the momentum dependence is the only observational consequence} of the loop corrections. We measure the mass $m^2 (p^2)$ at different energy scales specified by the {\em external momentum $p^2$} from the reference mass $m_h^2$. This external momentum here replaces what we usually call the renormalization scale $\mu^2$.

A timely example is the Higgs mass profile, dominated by loop corrections from the top quark \cite{Choi:2023cqs, Choi:2023mma}.

The divergence in the self-energy $\Sigma_l(p^2)$ is canceled by the counterterms of the same order so that each term in the bracket in (\ref{SlidingMass}) is as small as $\lambda^{l}$ with an order-one coefficient. Higher-order corrections give us a more precise dependence on momentum. Perturbation in QFT is well defined order by order, with finite and parametrically ordered coefficients, as justified in Section \ref{sec:regindep}.
We can generalize it to theories with multiple couplings, having a generalized self-coupling (\ref{prop}).

One immediate consequence of this is that if an amplitude does not depend on the external momentum, the corresponding amplitude is renormalized to zero. Otherwise, we have another problem of infinity arising from these amplitudes. This kind of amplitude always appears in the {\em internal} propagator and cannot be removed by the usual amputation done to the external propagator. 

This includes one-loop corrections to the Higgs mass-squared from the Higgs self-coupling and gauge bosons. 
For example, the leading correction comes from the one-loop, proportional to $\lambda,$
\be \label{Sigma1}
 - i \Sigma_1(p^2) =\frac12 \int \frac{d^4 k}{(2\pi)^4} (-6i \lambda) D(k^2).
 %=  \int \frac{d^4 k}{(2\pi)^4} \frac{3\lambda}{ k^2 - m^2} .
\ee
From the relations in (\ref{CTexp}), we see that $\delta_{Z1}=0$ and $(6\lambda) \delta_{m1} = -\Sigma_1$, so that the propagator (\ref{FeynProp}) remains the same at this order, from (\ref{denom}). 
Similar momentum-independent contributions arise from quartic interactions in non-Abelian gauge theories and from all types of tadpole diagrams.

%%%%%%%%%%%%%%%%%%%%%%%%%%%%%%%%%%%%%%%%%
\section{Renormalization without counterterms} \label{sec:woct}

We have another way to achieve renormalization without introducing counterterms.\footnote{The Wilsonian approach \cite{Wilson:1971bg} does not introduce counterterms and deals only with the bare parameters. The price to pay is that we instead have an unphysical cutoff momentum. However, we expect that the expansion parameters are not the bare ones but the physical ones, in the end \cite{Weinberg:1995mt, Choi:2023mma}, as we do here.}
This picture clarifies the relation between the bare and the renormalized masses. Moreover, we see that renormalization is not an artificial process of making a physical quantity finite, but the reorganization of the same quantity.

\subsection{Renormalization as rearrangement}

The free particle is described by the propagator, the inverse of the two-point vertex function,
\be \label{FeynPropB}
 D_B(p^2) = \frac{i}{p^2- m_B^2 },
\ee
encoding the information on the bare mass. It receives quantum corrections by the 1PI self-energy $\Sigma(p^2)$
\be \label{twoptfn}
  i D_B^{-1} (p^2) = p^2 - m_B^2 - \Sigma(p^2).
\ee
The mass, the location of the pole of the propagator (\ref{FeynPropB}), which can be probed by scattering experiments, is changed. We take a reference mass for the expansion as the pole mass $m_h$ defined as
\be \label{scpolemass}
  i D_B^{-1} (m_h^2) = m_h^2 - m_B^2 - \Sigma (m_h^2) \equiv 0.
\ee
We may also rewrite the two-point function (\ref{twoptfn}) in terms of the pole mass (\ref{scpolemass}),
\be
  i D_B^{-1} (p^2) = p^2 - m_h^2 - \Sigma(p^2) + \Sigma(m_h^2).
\ee
At this pole, the residue of the propagator is changed, which comes from the next leading order expansion of $\Sigma(p^2)$ from the reference mass
\be \label{proppole} \begin{split}
 p^2 \to m_h^2:  D_B (p^2) &= i \Big[ p^2-m_h^2 - (p^2-m_h^2) \frac{ d \Sigma}{d p^2}(m_h^2) + {\cal O}((p^2-m_h^2)^2)\Big]^{-1} \\
 & =  \frac{iZ}{p^2-m_h^2}+ (\text{terms regular at } p^2 = m_h^2).
\end{split}
\ee
We are free\footnote{From the LSZ reduction formula \cite{Weinberg:1995mt}.} to renormalize the field as in (\ref{fieldstr}), with the field strength $Z$ being
\be \label{wavefnren}
 Z^{-1} = 1- \frac{ d \Sigma}{d p^2}(m_h^2).
\ee
This means that the re-normalized field $\phi_r$ should be effectively the free field. 
The resulting propagator is the same as that in (\ref{FeynProp}) to ${\cal O}(p^2-m_h^2),$ justifying its use. Therefore, we calculate the amplitude, including the above $\Sigma(p^2)$, using the physical mass $m_h$. 

A scattering amplitude with the momentum $p^2$ will be described by the newly normalized propagator. From its denominator, we are naturally led to define the momentum-dependent  ``running mass''  (\ref{SlidingMass}), which can also be written as the Taylor expansion up to the second order in $p,$\footnote{Here and in what follows, $t^D_\mu$ denotes the Taylor expansion operator around the reference point $\mu$ up to order $D$ in the momentum, defined in (\ref{taylorop}) below.} from (\ref{proppole}),
\be \label{runningmass} 
  m^2(p^2) = m_h^2 + (1-t^2_{m_h}) \Sigma(p^2).
\ee
This structure has arisen from the field strength re-normalization (\ref{fieldstr}) with the value (\ref{wavefnren}) fit up to the quadratic order in $p$.

Of course, this description is equivalent to the previous one with counterterms if we absorb the counterterms for $\delta_m, \delta_Z$ in the mass and the propagator, not treating them as perturbative interactions. There is a one-to-one correspondence between the loop correction and the counterterm as in (\ref{CTexp}), so the separation between physical parameters and counterterms is unambiguous. 

Then, consider the one-loop correction to the quartic coupling $\lambda_B$
\be
 6\lambda_{(1)}(p_1,p_2,p_3,p_4) = 6\lambda_B + (6\lambda)^2 \Big( V(s) + V(t) + V(u) \Big),
\ee
where we define
\be \label{OLV}
 i V(p^2) \equiv \frac{1}{2} \int \frac{d^4 k}{(2\pi)^4} D(k^2) D((k+p)^2).
\ee
We have three channels described by the Mandelstam variables 
\be \label{ManRel}
s=(p_1+p_2)^2, \quad t= (p_1+p_3)^2,\quad  u=(p_1+p_4)^2,\quad  s+t+u=4m^2.
\ee
We compare the coupling with the experimental value via scattering
\be
 6\lambda =  6\lambda_{(1)}(\mu_1,\mu_2,\mu_3,\mu_4) = 6\lambda_B + (6\lambda)^2 \Big( V(s_0) + V(t_0) + V(u_0) \Big)
\ee
at a specific scale $(\mu_1,\mu_2,\mu_3,\mu_4)$ satisfying the relation (\ref{ManRel}),
\be
s_0 = (\mu_1+\mu_2)^2, \quad t_0 = (\mu_1+\mu_3)^2, \quad u_0 = (\mu_1+\mu_4)^2,\quad  s_0+t_0+u_0=4m^2.
\ee
This is the renormalization condition in the OS scheme. 
We may {\em rewrite} the corrected coupling in terms of this value
\be \label{OneComp} \begin{split}
 6\lambda_{(1)}(p_1,p_2,p_3,p_4) &= 6\lambda + (6\lambda)^2 \Big( V(s)  - V(s_0) + V(t) -  V(t_0) + V(u)  - V(u_0) \Big) \\
   & = 6\lambda + (1-t_\mu^0) (6\lambda)^2 \Big( V(s) + V(t) + V(u) \Big) \\
  & =6\lambda +  \sum_{(p^2,p_0^2)} \frac{36 \lambda^2}{32\pi^2} \int_0^1 dx \log \frac{m^2-x(1-x)p^2}{m^2-x(1-x)p_0^2},
\end{split}
\ee
where we have $(p^2,p_0^2) = (s,s_0), (t,t_0), (u,u_0)$. See the next section for the evaluation.

We see the renormalization is a {\em rewriting} of the same quantity: we expressed the coupling $ 6\lambda_{(1)}$ relative to the value $6\lambda$ at the reference scales $s_0,t_0,u_0$. It is {\em finite} renormalization. Further loop corrections are always {\em added as the difference} between the two quantities at these scales. 

%We have the full coupling 
%\be
% 6\lambda(p_1,p_2,p_3,p_4) = \lambda_B + Q (p_1,p_2,p_3,p_4),
%\ee
%which is compared to the experiment similarly
%\be
% \lambda = 6\lambda(\mu_1,\mu_2,\mu_3,\mu_4) = \lambda_B + Q (\mu_1,\mu_2,\mu_3,\mu_4),
%\ee
%so that
%\be \begin{split}
% 6\lambda(p_1,p_2,p_3,p_4) & = \lambda + Q (p_1,p_2,p_3,p_4) - Q (\mu_1,\mu_2,\mu_3,\mu_4) \\
% &= \lambda + (1-t_\mu^0) Q (p_1,p_2,p_3,p_4).
%\end{split}
%\ee

\subsection{Effective coupling and subdiagram renormalization} \label{secEff}

\begin{figure}[t!]
\begin{center}
\includegraphics[scale=0.7]{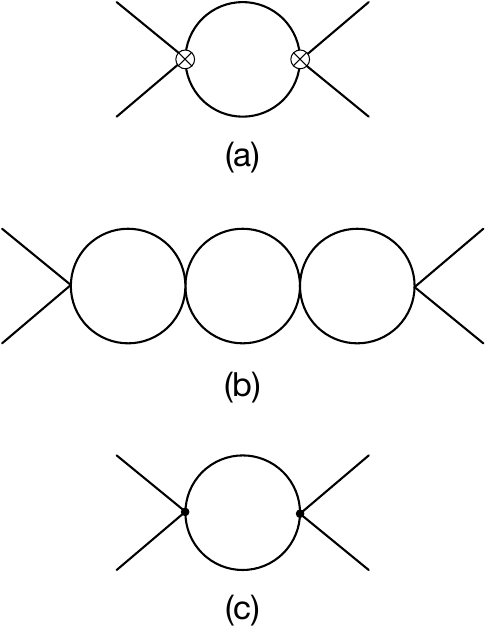}
\end{center}
\caption{Neither a loop-correction nor a counterterm is separately observable, but only the combination is. (a) The one-loop diagram involving the counterterm coupling $\delta_{\lambda 1}$ (denoted by crossed circles). The corresponding amplitude is divergent. A counter-counterterm to this may give rise to other divergences, e.g., having the same structure as Fig. (a) with the new counter-counterterm coupling. (b) {\em Special combination} with ordinary loop diagrams with physical couplings avoids the proliferation of counterterms. (c) If we always have the combination of the renormalized scale-dependent couplings (denoted by big dots), there is only a finite number of counterterms. \label{fig:delta1}}
\end{figure}

We may view a subamplitude as the correction to the tree-level coupling. Renormalizing this subamplitude works essentially the same way as performing a one-loop calculation. Consider the $s$-channel part of the one-loop counterterm
\be
 \frac{(6\lambda)^2}{24} \delta_{\lambda1} \phi^4 \supset -\frac{(6\lambda)^2}{24} V(s_0) \phi^4.
\ee
Since the counterterms provide interactions, we may consider a loop correction with them. It can also give rise to divergences, for example, from the amplitude in Fig. \ref{fig:delta1} (a)
\be
 \int  d^4 k  V(s_0) V((p+k)^2) V(s_0).
\ee 
Then, we worry that we need another quartic counter-counterterm to cancel the additional divergence. This may give rise to proliferating counter-counterterms, which is the criterion for non-renormalizability. However, our theory is renormalizable and should have a finite number of counterterms at a given order.

Only a particular combination can cancel the divergence arising from the counterterms. Noting that the renormalization adds the pair-subtracted loop $V(s)-V(s_0)$, we find that its main divergence can be canceled in the amplitude
\be
 \int d^4k_1 d^4 k_2 d^4 k_3 [V((p+k_1)^2) - V(s_0)] V((p+k_2)^2) [V((p+k_3)^2)-V(s_0)].
\ee
Its unrenormalized diagram is in Fig. \ref{fig:delta1} (b). This means that if we have the amplitude Fig. \ref{fig:delta1} (b), we always have Fig. \ref{fig:delta1} (a) and vice versa; whenever we have a counterterm parameter in the amplitude, we have an amplitude with this counterterm replaced with loop sub-amplitude. This also means that neither bare counterterm parameter is observable.

These two diagrams correspond to parts of the (subdiagram-)renormalized amplitude Fig. \ref{fig:delta1} (c),
\be \label{AmpCtCt} \begin{split}
 \lambda \int d^4k_1 d^4 k_2 d^4 k_3 &\Big(1+ 6\lambda [V((p+k_1)^2) - V(s_0)]\Big)V((p+k_2)^2) \\
 &\qquad \times \Big(1+ 6\lambda [V((p+k_3)^2) - V(s_0)]\Big).
\end{split} 
\ee
Here, the first and the third loops contain all the corrections up to one-loop order. They can be regarded as effective couplings and the corresponding subdiagram is denoted by big dots in Fig. \ref{fig:delta1} (c). 

We still have some divergence from some subdiagrams. We note that 
the amplitude (\ref{AmpCtCt}) is again a part of the completely renormalized quartic coupling
\be \label{V3s} \begin{split}
 \lambda \int d^4k_1  d^4 k_2 d^4 k_3 &\Big(1 + 6\lambda [V((p+k_1)^2) - V(s_0)]\Big)  \Big(1 + 6\lambda [V((p+k_2)^2) - V(s_0)]\Big) \\
 &\qquad \times \Big(1 + 6\lambda [V((p+k_3)^2) - V(s_0)]\Big).
\end{split}
\ee
This contains amplitudes from the tree level to the three-loop. In particular, the ${\cal O}(\lambda^4)$ counterterm should contain the counterterm for the three-loop part
\be
  \frac{(6\lambda)^4}{24} \delta_{\lambda4} \phi^4 \supset -\frac{(6\lambda)^4}{24} [V(s_0)]^3 \phi^4 + \text{(other  channels)}.
\ee
Of course, we have more diagrams from other channels \cite{Choi:2025tus}.

%%%%%%%%%%%%%%%%%%%%%%%%%%%%%%%%%%%%%%%%%%
\section{Regularization and scheme independence} \label{sec:regindep}

We emphasize that a physical quantity must be uniquely defined, independent of the choice of regularization or renormalization scheme. 

\subsection{Regularization independence}

Traditionally, we have handled divergent loop corrections and isolated the infinity as a limit of a parameter through regularization. No such problem arises, as done in the previous section, if we perform finite renormalization between two scales, which rewrites a physical quantity as the {\em difference} from the reference one. There is no reference to the UV scale; thus, the physical quantity has to be finite. 

Consider the quartic vertex correction (\ref{OLV}), which becomes the following after Feynman parametrization (e.g., \cite{Peskin:1995ev})
\be \label{QV}
 V(p^2) =  \frac{i}{2}  \int_0^1 dx  \int \frac{d^4 k}{(2\pi)^4} \frac{1}{\left( k^2 -\Delta(p^2) \right)^2},
\ee
where $\Delta(p^2) \equiv m^2-x(1-x)p^2$.
The divergence arises because the momentum integration extends to arbitrarily high scales. In the position space, it means that the product of two distributions at the same point is not well-defined \cite{Collins:1984xc}. This is due to the limitation of our only available calculational tool, the Lorentz-covariant continuum field theory. 

A simple (Euclideanized) momentum cutoff at $k_E^2 = k_0^2 +{\bf k}^2=\Lambda^2$ gives
\be \begin{split}
 i \int \frac{d^4 k}{(2\pi)^4} \frac{1}{\left( k^2 -\Delta \right)^2} 
 & = \frac{1}{16\pi^2}\left( - \log \frac{\Lambda^2 + \Delta}{\Delta} + \frac{\Lambda^2}{\Lambda^2+\Delta} \right) 
\\
 &= \frac{1}{16\pi^2}\left( \log \frac{\Delta}{\Lambda^2} + 1 + {\cal O}\left(\frac{\Delta}{\Lambda^2} \right)\right) \label{cutoff2}.
\end{split}
\ee
Dimensional regularization \cite{tHooft:1972tcz,Bollini:1972ui}, by taking the dimension $d \equiv 4 - \epsilon$, gives
\be
 i \int \frac{d^d k}{(2\pi)^d} \frac{1}{\left( k^2 -\Delta \right)^2} = \frac{1}{16\pi^2}\left(-\frac{2}{\epsilon} +\log \Delta +\gamma-\log(4\pi)+{\cal O}(\epsilon)\right), \label{DimRegLog}
\ee
where $\gamma$ is the Euler--Mascheroni constant. Comparing the two, the divergent parts correspond via $\log \Lambda^2 \leftrightarrow 2/\epsilon$, with regularization-dependent finite constants that will cancel in the renormalized expression. In general, we deal with the integral of the form \cite{Choi:2023mma, Weinberg:1995mt}
\be
 Q_0(\Delta) = i \int \frac{d^4 k}{(2\pi)^4} \frac{1}{\left( k^2 -\Delta \right)^2} =e \log \Delta + A_0,
\ee  
where the constant $A_0$ may contain regularized divergence and $e$ is an order one constant.
We fix the renormalization condition at a finite reference $\Delta_0 \equiv \Delta(m_h^2)$
\be
 Q_0(\Delta_0) =e \log \Delta_0 + A_0 
\ee
\be
 Q_0(\Delta) - Q_0(\Delta_0) =e \log \frac{\Delta}{\Delta_0}.
\ee

The lesson from the last section is that the physical quantity after renormalization is expressed as the difference between nearly identical quantities at different scales. In the simplest case of the marginal quantity, they are the same.
{\em Any} of the regularizations, (\ref{cutoff2}) or (\ref{DimRegLog}), gives the same renormalized quartic vertex in (\ref{OneComp}).
Any renormalization scheme thus requires a reference scale \cite{Georgi:1993mps}. 
Moreover, we see that the difference is {\em insensitive} to the high-momentum modes of $k$, independent of $\Lambda$ or $1/\epsilon.$ We will further see that the renormalization does {\em not} care about the {\em divergence} of the loop amplitudes. In general, the two are not exactly the same, but they are sufficiently close to remove the divergences only. 

This natural cancellation behavior is systematically rephrased as the Bogoliubov--Parasiuk--Hepp--Zimmermann (BPHZ) scheme \cite{Bogoliubov:1957gp,Hepp:1966eg,Zimmermann:1967,Zimmermann:1969jj,Lowenstein:1975ps,Lowenstein:1975rg}; see also \cite{Blaschke:2013cba,BS}. 
The renormalized quantity $\Gamma(p^2)$ is defined as the momentum integral up to a boundary value $\Gamma (\mu^2)$ at a renormalization point
\be\label{BPHZQ}
 \Gamma(p^2) = \Gamma (\mu^2) + (1-t_\mu^D) \int I(p^2),
\ee
where $D$ is the superficial degree of divergence \cite{Weinberg:1959nj}. The integration for the Feynman diagram is done over the loop momenta. 
As we have seen in the mass, we include the Taylor expansion operator $t_\mu^D$ up to $D$ differentiations in $p$ around the renormalization point \cite{Dyson:1949ha, BS55}
\be \label{taylorop}
 t^D_\mu f(p) \equiv \sum_{n=0}^D \frac{1}{n!} (p_{i_1}^{\alpha_1}-\mu_{i_1}^{\alpha_1})\dots  (p_{i_n}^{\alpha_n}-\mu_{i_n}^{\alpha_n})\frac{\partial^n f}{\partial p_{i_1}^{\alpha_1} \dots \partial p_{i_n}^{\alpha_n}} (\mu),
\ee
because they all interfere in the observed scattering, forming inseparable observables. The Taylor expansion arises from two equivalent conditions. The first is the counterterm adjustment shown in (\ref{ZRenCond}). More fundamentally, it is required so that renormalized correlation functions maintain the same analytic behavior as their tree-level counterparts, as expressed in (\ref{wavefnren}). The functions appearing in this paper are all invariant under $p \to -p$ so that they are functions of $p^2$. Generalization to multivariable cases is straightforward.
The above removes the main divergence, assuming that possible subdiagram divergences are canceled by the BPHZ prescription \cite{Bogoliubov:1957gp, Hepp:1966eg, Zimmermann:1967, Zimmermann:1969jj, Lowenstein:1975ps, Lowenstein:1975rg}.

The quartic vertex correction (\ref{QV}) has $D=0$, and the Taylor expansion has only the constant term.
The self-energy (\ref{selfE}) has $D=2$ and contains quadratic divergence in four dimensions. 
There is a natural cancellation of the parameter $\Lambda$ or $\epsilon$ as well as the scheme-dependent constants \cite{Choi:2023cqs}. This is because the loop correction of the mass-dimension-two quantity is spanned by the previous integral (\ref{cutoff2}) or (\ref{DimRegLog}) and the following:
\begin{align}\begin{split}
  i \int \frac{d^4 k}{(2\pi)^4} \frac{1}{k^2 -\Delta }
   %=\frac{ 1}{16\pi^2} \int_0^{\Lambda^2}d (k^2) \frac{\ell^4}{(\ell^2-\Delta)^2} 
   &=\frac{1}{16\pi^2} \left( \Lambda^2 - \Delta \log \frac{\Lambda^2+\Delta}{ \Delta} \right)\\
   &= \frac{1}{16\pi^2} \left( \Lambda^2 - \Delta \log \frac{\Lambda^2}{ \Delta}+{\cal O}\left(\frac{\Delta^2}{\Lambda^2} \right) \right), 
   \end{split} \label{cutoff1} \\
 i \int \frac{d^d k}{(2\pi)^d} \frac{1}{ k^2 -\Delta }  &=\frac{1}{16\pi^2}\left(-\frac{2}{\epsilon}\Delta + \Delta \log \Delta + \gamma \Delta - \Delta \log(4\pi) - \Delta +{\cal O}(\epsilon) \right). \label{DimRegQuad} 
\end{align}
The substantial difference is that dimensional regularization neglects the quadratic divergence. The scheme dependence is the linear terms in $\Delta$. 

This is no problem since each of them is treated as a constant and {\em canceled in the physical quantity} (\ref{runningmass}) \cite{Choi:2023cqs, Choi:2023mma}. We may obtain (\ref{cutoff1}) or (\ref{DimRegQuad}) from the $Q_0(\Delta)$ in (\ref{cutoff2}) using {\em the same regularization}, 
\be \label{I2}
 Q_2(\Delta) = \int d \Delta Q_0 (\Delta) =e \Delta \log \Delta + (A_0 -e )\Delta + A_2,
\ee
where $A_2$ is again the integration constant that may contain the isolated infinity.
Although $Q_2(\Delta) - Q_2(\Delta_0)$ removes the regularized divergence $A_2$, it does not remove $A_0 -e $. Nevertheless, we see that the Taylor expansion completely removes $A_0$ and $A_2$
\be \label{renI2}
 Q_2(\Delta) - Q_2(\Delta_0) - (\Delta-\Delta_0) \frac{d Q_2}{d \Delta} (\Delta_0) = e\Delta \log \frac{\Delta}{\Delta_0} -e \Delta +e \Delta_0. 
\ee 
This is so because the differentiation $dQ_2/d\Delta$ contains the newly entered constant $A_0$. Then $A_0 (\Delta-\Delta_0)$ contained in $Q_2(\Delta)-Q_2(\Delta_0)$ is removed by $(\Delta-\Delta_0) dQ_2/d\Delta$. We see that any linear terms in $\Delta$ vanish in this process, and the linear term in (\ref{renI2}) comes from the logarithm in (\ref{I2}), as the coefficients show. 

Since $\Delta$ is linear in $p^2$, the relation holds
\be \label{Deltaprel}
  \Delta(p^2)- \Delta(m_h^2)-  (p^2-m_h^2)  \frac{d \Delta}{dp^2}(m_h^2) = 0,
\ee
translating the dependence of $\Delta$ into that of $p^2$. 
Using this, we see that (\ref{renI2}) now becomes
\be \begin{split}
 (1-t^2_{m_h}) Q_2(p^2) &= Q_2(p^2) - Q_2(m_h^2) - (p^2-m_h^2) \frac{dQ_2}{dp^2}(m_h^2) \\
 & = e \Delta(p^2) \log \frac{\Delta(p^2)}{\Delta(m_h^2)}- e \Delta(p^2)+e \Delta(m_h^2).
\end{split}
\ee
This is the renormalized mass (\ref{runningmass}), which is finite and independent of regularization.\footnote{When dimensional regularization is combined with BPHZ subtraction at a physical reference, the scale dependence of the observable is carried by the external momentum $p^2$ rather than by an arbitrary scheme parameter $\mu$, eliminating the need for explicit matching between separate effective theories \cite{Georgi:1993mps, Georgi:1976vf}.}

We may generalize the above.
The loop integrals appearing at higher orders reduce, after the Feynman parametrization (see (\ref{compden}) below), to functions of $\Delta$ of mass dimension $2n$, whose dependence on $\Delta$ is spanned by the iterated indefinite integrals of $Q_0(\Delta)$,
\be
 Q_{2n}(\Delta) \propto  \left(\int d \Delta \right)^n Q_0 (\Delta) = \frac{e}{n!} \Delta^n \log \Delta + P_n(\Delta),
\ee
where $P_n(\Delta)$ is a polynomial of degree $n$ whose coefficients may contain the regularized divergences. For $n=1$, this reproduces (\ref{I2}).
Note that the expressions with $(n+1)$ are obtained by an indefinite integration of those with $n$ with respect to $\Delta$. As long as the relation (\ref{Deltaprel}) holds, this is essentially an integration with respect to $p^2$.
Therefore, we can show that the polynomials in $\Delta$ cancel in the physical quantity $\Gamma(p^2)$ in (\ref{BPHZQ}) and the physical quantity contains terms involving only
\be \label{deltadelta} 
(1-t^{2n}_{m_h})Q_{2n}(\Delta) \propto \Delta^{n} \log \frac{\Delta(m_h^2)}{\Delta(p^2)} + P(\Delta(p^2)),
\ee
where a polynomial $P(\Delta)$ in $\Delta$ is completely generated from the integration of $Q_0 (\Delta) = \log \Delta$. This quantity (\ref{deltadelta}) is free of divergence and {\em independent of regularization.} 
Regardless of regularization, Zimmermann showed that the integral (\ref{BPHZQ}) is convergent by slightly modifying the Feynman prescription, even if the integral is Lorentzian  \cite{Zimmermann:1969jj}.

The above shows that the momentum cutoff $\Lambda$ in the integral has nothing to do with the hierarchy. What we require is that the loop corrected quantity we calculate should be sufficiently close to the subtracted terms (the counterterms or Taylor expansions). Restoring the regulator as a second argument, and writing $\Sigma^{\text{ren}}(p^2)$ without it for the value after the regulator is removed, we have from (\ref{cutoff1}) \cite{Choi:2023cqs}
\be \label{CutoffAcc}
 \Sigma^{\text{ren}}(p^2;\Lambda) = \Sigma^{\text{ren}}(p^2)
 + {\cal O} \left( \frac{(p^2-m_h^2)^2}{\Lambda^2} \right),
\ee
since the terms of (\ref{cutoff1}) that survive the expansion in $\Delta/\Lambda^2$ are quadratic in $\Delta$ and hence in $p^2$. Due to the natural pair-subtraction property, sufficient high-momentum contributions cancel and $\Lambda$ does not appear in the (super-)renormalizable parameters. We may allow $\Lambda^2$ not so far from the scale of the probe $p^2$.

The same holds in dimensional regularization, where the quadratic divergence is absent from the start. The integral in (\ref{DimRegQuad}) is exactly
\be \label{DimRegExact}
 Q_2(\Delta;\epsilon) = \frac{\Delta}{16\pi^2}
  \left( \frac{4\pi}{\Delta} \right)^{\epsilon/2}
  \Gamma\!\left( -1 + \frac{\epsilon}{2} \right),
\ee
whose expansion one order beyond (\ref{DimRegQuad}) reads
\be \label{DimRegOrdEps}
 Q_2(\Delta;\epsilon) = \frac{1}{16\pi^2} \left[
  -\frac{2\Delta}{\epsilon} + \Delta \log \Delta
  + (\gamma - \log 4\pi - 1) \Delta
  - \frac{\epsilon}{4} \Delta \left(
    \log^2 \frac{4\pi}{\Delta} + 2(1-\gamma) \log \frac{4\pi}{\Delta}
    + \text{const} \right) \right] + {\cal O}(\epsilon^2).
\ee
The pole and the scheme constants multiply $\Delta$ linearly, so they are removed by (\ref{Deltaprel}) exactly as the constants $A_0$ and $A_2$ were in (\ref{renI2}). What is new at order $\epsilon$ is not linear in $\Delta$, so it survives the subtraction, but it carries an explicit factor of $\epsilon$. Applying $1-t^2_{m_h}$ to (\ref{DimRegOrdEps}) and using (\ref{deltadelta}) for the $\Delta \log \Delta$ and $\Delta \log^2 \Delta$ terms,
\be \label{DimRegAcc}
 \Sigma^{\text{ren}}(p^2;\epsilon) = \Sigma^{\text{ren}}(p^2)
 + {\cal O}\!\left( \epsilon \, \frac{(p^2-m_h^2)^2}{\Delta(m_h^2)} \right),
\ee
up to logarithms. In both regularizations the renormalized self-energy differs from its regulator-removed value only at second order in $p^2-m_h^2$, and the difference is controlled by the regulator alone. This is what we mean by saying that the physical mass is independent of regularization.

Now, we see that the theory admits a well-defined perturbative expansion. In $\phi^4$-theory, while each $\Sigma(p^2)$ and $\lambda_{(1)}(p)$ are divergent, their renormalized counterparts $(1-t^2_{m_h}) \Sigma(p^2)$ and $(1-t^0_\mu) \lambda_{(1)}(p)$ are parametrically small and admit perturbative expansions. 

The extreme case is the massless $\phi^4$ theory. The one-loop self-energy is a tadpole,
\be
 \Sigma_1 = \frac{6 \lambda}{2} \int \frac{d^4 k_E}{(2\pi)^4} \frac{1}{k_E^2}
          = \frac{3 \lambda \Lambda^2}{16\pi^2} ,
\ee
which, taken at face value, would generate a large mass in a theory that is classically scale invariant and has none. It carries no momentum dependence, so $(1-t_0^2)\Sigma_1 = 0$ identically, and in dimensional regularization the integral is scaleless and vanishes outright. Beyond one loop, the subtraction point $p^2=0$ is itself infrared singular and the
prescription must be modified. Lowenstein and Zimmermann \cite{Lowenstein:1975ps} showed that this can be done by introducing an auxiliary mass and separate ultraviolet and infrared subtraction degrees, so that every diagram of the massless model contributes to the Green functions in a way that is convergent in both regimes and the theory remains massless order by order. A quadratic dependence on $\Lambda$ is therefore not by itself evidence that the scalar mass is sensitive to short-distance physics. This also separates the two questions that are often merged. 
A dependence on the regulator $\Lambda$ is removable by construction. A dependence on a physical heavy mass $M$ is not, and whether it occurs is a question about the theory rather than about the scheme. We take up the latter in Section \ref{sec:decoup}.

%%%%%%%%%%%%%%%%%%%%%%%
\subsection{Scheme equivalence}

Regarding the renormalization as a finite ``scale transformation,'' the only remaining freedom is the fixing of the reference quantity. The choice of a different renormalization scheme sets another reference point, which can be altered again by finite renormalization. With this, we show that the OS scheme is equivalent to the BPHZ scheme. 

Consider a bare coupling before the field strength renormalization, receiving loop corrections
\be \label{gammaatp}
 \langle \phi_1(p_1) \dots \phi_n (p_n) \rangle_{\rm 1PI} =  \Gamma^{(n)}(p_1,\dots p_n) = g_B+ \int I(p_1,\dots p_n).
\ee 
In this paper, it suffices to consider the case $n=2$, in which $\Gamma^{(2)}(p^2) \equiv  \Gamma(p^2)$ depends on $p^2$ and the generalization is straightforward.
The renormalization condition fixes the bare parameter at a reference scale $p^2 = \mu^2$
\be \label{gammaatmu}
 \Gamma(\mu^2) = g_B+ \int I(\mu^2).
\ee
Taylor expansion removes the bare coupling and makes it a relative quantity
\be \begin{split}
 \Gamma(p^2) &=  \Gamma(\mu^2) +  \int I(p^2) - \int I(\mu^2) \\
 &=  \Gamma(\mu^2) + (1-t^D_\mu)  \int I(p^2) .
\end{split}
\ee
The field strength re-normalization gives the relation (\ref{BPHZQ}), adding more terms that are the Taylor expansion.

Thus, BPHZ uses the finite quantity
\be \label{BPHZdef}
 \Gamma (p^2) =  \Gamma(\mu^2) +\int \big(1-t_\mu^D \big) I(p^2),
\ee
which is regularization-independent by default (if the exchange is permissible). If we work directly with the subtracted pair rather than with the individual divergent terms, we do not need to modify short-distance physics by introducing fields with wrong spin statistics or heavy vector bosons \cite{Georgi:1993mps}. There is no arbitrariness in the quantity $\Gamma(p^2)$ except for the integration constants, which are to be completely fixed by the experiments. 

The original BPHZ prescription uses the zero momentum as the reference point. However, we can move the reference point by finite renormalization \cite{Hepp:1966eg} 
\be \label{finiteren} \begin{split}
 \Gamma (p^2) &=  \Gamma(\mu^2) +\int \big( 1-t^D_0 + t^D_0-t^D_\mu \big) I(p^2) \\
 & = \Gamma(0) + \int \big( 1-t_0^D \big) I(p^2) + {\cal O}\big((p^2-\mu^2)^2\big).
\end{split}
\ee
Since there is no regularization dependence, the only scheme dependence is the reference parameter at different scales. Up to this point, any renormalization scheme yields the same result. In fact, all the renormalizations we use are finite, relating the scale-dependent quantity $\Gamma(p^2)$ as the relative value at a reference scale $\Gamma(0)$ or $\Gamma(\mu^2)$, except for relating the bare and the physical parameters.

In the OS scheme, $\Gamma(\mu^2)$ is the experimentally observed value via scattering experiments done at $p^2=\mu^2$ so that further corrections in (\ref{BPHZdef}) do not modify it. If the scheme is instead defined such that the experimental observable is partly contained in the $\int \big(1-t_\mu^D \big) I(p^2)$, higher-order corrections will make the total value arbitrarily close to the observed value.

Renormalization is {\em not about removing or absorbing the divergences} but about comparing one physical quantity at different scales.
In the above example, the effective mass $m^2(p^2)$, in the end, is given by the solution to the renormalization group equation for the running physical parameter $\Gamma(p^2)$ specified by $I(p^2)$,
\be \label{RGE}
 p^2 \frac{d \Gamma(p^2)}{d p^2} - p^2 \frac{d }{d p^2} \int (1-t^D_\mu) I(p^2) = 0,
\ee 
with the initial condition set at a reference scale $m_h^2$. It is easy to see that, for the quartic coupling at one loop, this reproduces the Callan--Symanzik equation with the additional terms from the field-strength renormalization. For instance, the one-loop corrected quartic coupling $\lambda_{(1)}$ in (\ref{OneComp}) is the solution to the partial differential equation
\be
s\frac{\partial \,(6\lambda_{(1)})(s,t,u)}{\partial s} 
=   \frac{36\lambda^2}{32 \pi^2} \int_0^1 dx \frac{-x(1-x)s}{m^2-x(1-x)s}  \underset{m^2\to0}{\to} \frac{36\lambda^2}{32 \pi^2},
\ee
and similar for the other channels $t,u$, 
with the boundary condition being the renormalization condition.
Likewise, we see that the renormalized coupling in (\ref{BPHZdef}) is the solution to the equation (\ref{RGE}). The renormalization-group flow is finite, ranging between the scales $m$ to $\sqrt{p^2}$, and therefore, the physical quantity $\Gamma(p^2)$ does not see a high-energy contribution (see also \cite{Wetterich:1983bi, Mooij:2021ojy}).\footnote{The Wilsonian approach more clearly shows this \cite{Choi:2023mma}. In fact, the cutoff $\Lambda$ is human-made, which is overlooked in the Wilsonian renormalization that makes use of it explicitly. Eventually, it must be replaced by another dimensionful parameter, e.g., the mass $M$, of the fields involved in the loop provided by Nature  \cite{Choi:2023mma, Weinberg:1995mt}.}

In this regard, we must faithfully follow the renormalization procedure. If we remove only the infinite parts, as in minimal subtraction, the expansion parameters are no longer the physical ones and the decoupling of heavy fields is not manifest; the finite parts must then be restored by explicit matching at each threshold. Keeping the complete counterterms expresses the result directly in terms of the physical parameters.

Specifying the renormalization conditions at an agreed scale completely fixes the physical parameters. 
Since this is the only {\em observable combination,} it is natural to define this as the {\em effective coupling,} which is scale-dependent and fits the picture we had in Section \ref{secEff}.

%%%%%%%%%%%%%%%%%%%%%%%%%%%%%%%%%%%%
\section{The technical hierarchy problem}

Recall the usage fixed in the introduction. The dependence on the regulator $\Lambda$ is a statement about the parametrization and is disposed of in the present section. The term {\em technical hierarchy problem} is reserved for the dependence of the observable scalar mass on a physical heavy threshold $M$, which Section \ref{sec:decoup} answers.

\subsection{Old formulation}

Understanding the meaning of the bare mass is our starting point.
If one regards the bare mass $m_B^2$ in (\ref{phi4Lag}) as ``God-given'' from the beginning, it looks miraculous to have the rest of the corrections canceled to perfect accuracy to yield the observed small mass
\be \label{renmass}
m^2(p^2) = m_B^2 + \sum_{\{l_i\}}  \left[  \Sigma_{l_1,l_2,\dots,l_n}(p^2) - (p^2 -m_h^2) \frac{ d \Sigma_{l_1,l_2,\dots,l_n}}{d p^2} (m_h^2) \right],
\ee
using the notation in (\ref{prop}). We would fine-tune and force the quantum-corrected mass to be finite at a given order. However, higher-order corrections inevitably disrupt this fine-tuning, causing the sum to diverge.
There is, in general, more than one coupling, so some coupling we have yet to consider would also ruin the finite mass. In the Standard Model, we have three gauge and Yukawa couplings, in addition to the quartic self-coupling \cite{Kraus:1997bi}. Every loop correction from them has divergence. So, we would need more delicate relations between counterterms. 

Traditionally, the hierarchy problem has been why the correction $\Sigma(p^2)$ must be small to match the observed small value of $m^2(p^2)$, assuming small $m_B^2$ for some reason.
Its understanding has been guided by the naturalness criterion of 't Hooft \cite{tHooft:1979rat}.\footnote{If we set the parameter (here, the bare mass) to zero, there may emerge a symmetry. Then, this parameter serves as the order parameter of the symmetry breaking. This means every loop correction is proportional to this parameter, or the renormalization is multiplicative. This explains the smallness of the whole parameter: if the original parameter is small, all the corrections are proportional to it, and the total coupling becomes parametrically small. Examples include gauge symmetry and chiral symmetry. Due to the latter, for instance, a fermion mass correction is multiplicative and proportional to the mass itself, reducing the actual mass dimension of the self-energy, which has milder divergence. However, it is well known that scalar mass is not protected by symmetry and is not multiplicatively renormalizable.} 
Most importantly, we do not need to explain the smallness of the $\Sigma (p^2)$ part separately, because $\Sigma (p^2)$ and the bare parameter are {\em never separable}. What needs to be small is the {\em combination} (\ref{renmass}) \cite{Choi:2023cqs, Aoki:2012xs}.

\subsection{New formulation} \label{sec:newform} 

Then, how does the cancellation in (\ref{renmass}) take place? 
We use the term ``cancellation'' when comparing two independently known quantities. However, the bare parameter is merely what we define within our theory, and its redefinition should not be called cancellation.
Remember that our theory is an effective field theory, and the parameter $m_B^2$ in the Lagrangian simply parametrizes our ignorance about physics beyond a certain scale. It is to be matched with the counterpart in the UV completion or simply fitted by experiments (see, e.g., \cite{Cohen:2019wxr, Brivio:2017vri}). 

Although the bare parameter defines the theory, it is not observable. What is observable is the totally corrected combination (\ref{renmass}), which can be determined without reference to the bare parameter. 
We may {\em define} the theory using the pole mass $m_h$ that is measurable, then the bare mass is {\em derived} from the relation (\ref{baremassagain}) \cite{Coleman:2018mew}\footnote{An earlier formulation of this idea is mentioned, without specific attribution, in \cite{Bardeen:1995kv}.},
\be 
 m_B^2  = m_h^2 - \sum_{\{l_i\}} \Sigma_{l_1,l_2,\dots,l_n}(m_h^2),
\ee
which is nothing but the separation between the physical and the counterterm parameters. Then, the same mass (\ref{renmass}) is expressed as
\be \label{SMQ}
\begin{split}
  m^2(p^2) &= m_h^2 +  \sum_{\{l_i\}} \left[ \Sigma_{l_1,l_2,\dots,l_n}(p^2) - \Sigma_{l_1,l_2,\dots,l_n}(m_h^2) - (p^2 -m_h^2) \frac{ d \Sigma_{l_1,l_2,\dots,l_n}}{d p^2} (m_h^2) \right] \\
  & \equiv m_h^2 + \sum_{\{l_i\}} (1-t_{m_h}^2)   \Sigma_{l_1,l_2,\dots,l_n}(p^2) \\
   & \equiv m_h^2 +\sum_{\{l_i\}}  \Sigma^{\text{ren}}_{l_1,l_2,\dots,l_n}(p^2).
\end{split}
\ee
(Precisely speaking, the whole two-point vertex function or the propagator (\ref{gamma2}) is observable.)

Thus, the two views are equivalent.
For instance, we may define {\em the same} $\phi^4$-theory by either Lagrangian (\ref{phi4Lag}) or (\ref{altdef}). In the latter, the pole mass $m_h^2$ is fixed and calculating higher-order loop corrections ``reveals'' the bare mass more precisely. Conversely, in the original formulation (\ref{phi4Lag}), we regard the bare mass $ m_B^2$ fixed and the calculation more precisely corrects the (constant part) of the physical parameter. 

This reformulation leads to several important consequences.
Firstly, since the mass is always expressed in this combination, the cancellation of divergence is built in, as we have seen in the previous section. Being the subtracted combination, we understand why the loop corrections are canceled to be finite order by order, independent of the regularized parameters such as $\Lambda$ and $1/\epsilon$.
Each term in the square bracket is interpreted as the perturbative expansion of the mass, giving the relation (\ref{totmassub}) in coupling $g_1,g_2,\dots g_n$. 
%As long as there is no convergence problem, it is not difficult to convince ourselves that each correction becomes smaller and the perturbation works. 
Higher-order terms and/or loops in different couplings are parametrically small, allowing us to approximate and calculate them to the desired accuracy. We need Dirac's naturalness only: all the parameters appearing in the loop correction are of order one \cite{Dirac:1937ti}.

Secondly, the renormalized correction may still depend on the physical mass $M$ of a UV field. If such a field is in the loop, dimensional analysis shows that the renormalized correction should be
\be \label{UVsensitivity}
\Sigma^{\text{ren}}_{l_1,l_2,\dots,l_n}(p^2)= {\cal O} (M^2 \log M^2).
\ee 
The original formulation by Gildener essentially addressed this problem concerning stability against higher-order loop corrections \cite{Gildener:1976ai, Gildener:1979dd}. A loophole of the argument is that there are other dimensionful parameters so that combinations like $m_h^4/M^2,m_h^2 p^2/M^2$ or $p^4/M^2$ are also possible. In the next section, we show that the full calculation as above indeed gives these suppressed results.

\subsection{Relation to the Wilsonian formulation}

The statement that the quadratic divergence is not an observable has been made before in the Wilsonian language, and it is worth marking what is shared and what is not. In that formulation, the scalar mass is a relevant coordinate on theory space and the quadratic divergence is a shift of that coordinate rather than a physical effect, so that the meaningful quantity is the distance from the critical surface and not the value of the bare parameter at any particular cutoff \cite{Wetterich:1983bi, Aoki:2012xs, Yamada:2020cvt,Branchina:2022gll}. We agree with this reading, and the reformulation of Section \ref{sec:newform} is its counterpart in the language of subtractions. The bare mass and the loop correction are two ways of writing the same coordinate, and only the combination (\ref{SMQ}) is measured.

Two differences are worth recording. First, the Wilsonian argument is carried out with an explicit cutoff, and the object under discussion is the flow of a bare parameter with that cutoff. The formulation here introduces no cutoff at all. The subtracted combination is finite by construction \cite{Zimmermann:1969jj}, so the statement does not depend on what the regulator is or on whether one exists, and the conclusion is the same in dimensional regularization, where no quadratic divergence appears in the first place.

Second, and more important for what follows, the Wilsonian discussions settle the status of $\Lambda$ but not that of a physical heavy mass $M$. Once a heavy field is present in the loop, the correction to the scalar mass can depend on $M$, and no argument about the meaning of the regulator removes that dependence. This is the question we take up in Section \ref{sec:decoup}, where the two formulations part ways.

%%%%%%%%%%%%%%%%%%%%%%%%%%%%%%%%%%%%%%%%%%%%%%
\section{Decoupling in the scalar mass} \label{sec:decoup}

Finally, we propose that decoupling resolves the hierarchy problem associated with loop corrections. So far, we have seen that the loop corrections are perturbative and cutoff-independent. It remains to be shown that any of them involving heavy fields are suppressed.\footnote{Kazama and Yao \cite{Kazama:1981fx} first connected decoupling to the hierarchy problem. Their analysis of decoupling left residual ``fully reduced'' terms. Our work treats the full renormalization, including the counterterms, giving suppression without residual terms.}

The Appelquist--Carazzone decoupling theorem states that a 1PI amplitude involving a heavy field of mass $M$ in the loop is suppressed as powers of $p^2/M^2$  \cite{Appelquist:1974tg}. 
It has been shown that the heavy fields decouple in the strictly renormalizable (marginal) quantities, so that we do not need to care about heavy fields in the low-energy limit. The scalar mass-squared is a relevant parameter and is treated separately in the original statement of the theorem, which holds up to the renormalization of the masses of the light theory. What we do here is to make the statement explicit for the scalar mass and to show that, once the observable is correctly identified, nothing is left outside it. Throughout we assume that the couplings are held fixed as $M \to \infty$, which is the condition that the vertices themselves do not grow with the heavy mass \cite{Appelquist:1974tg, Schimert:1983fb}.

Let us suppose that, in addition to the ``Higgs'' scalar $\phi$ described in (\ref{phi4Lag}), there are a scalar $X$ and a Dirac field $\psi$ at the UV having interactions
\be \label{UVLag}
\Delta {\cal L} = - \frac12 M_s^2 X^2 -\frac14 \kappa \phi^2 X^2 - M_f \overline \psi \psi - y \phi \overline \psi \psi +\dots .
\ee
The UV scale is characterized by huge values of $M_s$ and $M_f$. (The masses in (\ref{UVLag}) are bare; however, we assume that they are to be renormalized in the same sense as before, and for brevity, we denote them in the same notation without confusion.) The decoupling limit is $M_s, M_f\to \infty$ or $M_s^2, M_f^2 \gg m_h^2, p^2$ to be precise.
Throughout, the decoupling limit is taken {\em with the couplings held fixed.} This hypothesis is essential: if a heavy mass originates from spontaneous symmetry breaking, so that a coupling grows with the mass, as $y = M_f/v$ for a chiral fermion acquiring its mass from a vacuum expectation value $v$, the suppression derived below is compensated by the growing coupling, $y^2(p^2-m_h^2)^2/M_f^2 = (p^2-m_h^2)^2/v^2$, and the field does not decouple, in accordance with the well-known non-decoupling effects of heavy chiral fermions. The masses $M_s$ and $M_f$ in (\ref{UVLag}) are explicit parameters independent of the couplings, so the hypothesis is satisfied.

Importantly, the main result presented in Sec. \ref{subsec:gendecoup} considers the most general, all-order loop correction, including the interaction of fields of arbitrary spins.

\subsection{Decoupling of a fermion at one loop}

\begin{figure}[t]
\begin{center}
\includegraphics[scale=0.7]{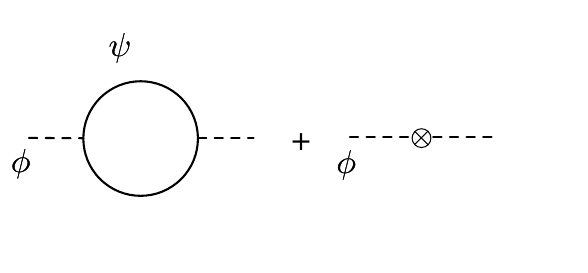}
\end{center}
\caption{The one-loop self-energy of a scalar $\phi$ involving a UV fermion $\psi$ of the mass $M$ and the counterterm contribution. There are two counterterms in $p^0,p^2$ with the external momentum $p$, although we draw one collectively. Although each does not vanish in the decoupling limit $M \to \infty$, the sum does. \label{fig:psioneloop} }
\end{figure}

As a warm-up, we first show that the one-loop correction by the fermion exhibits decoupling \cite{Choi:2023mma}. The amplitude comes from the Yukawa interaction in (\ref{UVLag}) 
 \be 
-i \Sigma_2(p^2) = - 2 y^2 N_c \int_0^1 dx \int \frac{d^4 \ell}{(2\pi)^4} \frac{\ell^2+\Delta(p^2)}{(\ell^2-\Delta(p^2))^2}, \ee
or
\be \label{oneloopfermi}
\Sigma_2(p^2) =- \frac{y^2 N_c}{8\pi^2} \int_0^1 dx \left[ \Lambda^2 - 3\Delta(p^2) \log \frac{\Lambda^2}{\Delta(p^2)} + 2\Delta (p^2)\right],
\ee
where the subscript in $\Sigma_2(p^2)$ denotes the dependence on the Yukawa coupling and we define 
\be \Delta(p^2) \equiv M_f^2-x(1-x)p^2.
\ee
Here, $N_c$ is the number of colors accounting for the multiplicity of the fermion \cite{Feng:2013pwa}. Since the mass of the fermion is not given by the electroweak symmetry breaking, $M_f/y$ is not the electroweak vacuum expectation value. 

We introduced a momentum cutoff $\Lambda$.
Although $\Sigma_2(p^2)$ diverges as $\Lambda^2$, the combination (\ref{SlidingMass})
\be \label{oneloopres}
 \Sigma_2^{\text{ren}}(p^2) = -\frac{y^2 N_c}{16\pi^2} \left[p^2-m_h^2+ 6 \int_0^1 dx\, \Delta(p^2) \log \frac{\Delta(p^2)} {\Delta(m_h^2)}\right] ,
\ee
and hence the running physical mass $m^2(p^2)$ in (\ref{SlidingMass}) converges, that is, free of $\Lambda$. The same result is obtained, independent of the choice of regularization.

Since $\Delta(p^2) = {\cal O}(M_f^2),$ at first sight, it seems to show a typical UV sensitivity (\ref{UVsensitivity}). However, in the decoupling limit $M_f \to \infty$, we have
\be \label{logexp}
\log \Delta(p^2) = \log (M_f^2) -\frac{x(1-x) p^2}{M_f^2} -\frac{x^2(1-x)^2 p^4}{2 M_f^4} +{\cal O}\left( \frac{p^6}{M_f^6} \right),
\ee
so that
\be 
 \Sigma_2^{\text{ren}}(p^2) = -\frac{y^2 N_c (p^2-m_h^2 )^2}{16\pi^2 M_f^2} \left[ \frac{1}{10} + \frac{p^2+2m_h^2}{140 M_f^2} \right] +{\cal O}\left( \frac{p^8}{M_f^6} \right).
\ee
It is suppressed as $(p^2-m_h^2)^2/M_f^2$ and vanishes in the limit $M_f\to \infty.$ The cancellation of $\Lambda$ is not a tuning because the $\Sigma_2^{\text{ren}}(p^2)$ in (\ref{oneloopres}) and hence $m^2(p^2)$ in (\ref{SlidingMass}) is the difference of the same functions with different arguments. Alternatively, an equivalent definition of $m^2(p^2)$ in the way in (\ref{BPHZdef}) contains no cutoff $\Lambda$ from the beginning.

\subsection{General decoupling in the scalar mass} \label{subsec:gendecoup}

Now, we present the general formula. We note that a higher order amplitude should also be the function of the similar arguments $\Delta(p^2)$ and $\Delta(m_h^2)$ of the logarithm in (\ref{oneloopres}), as the integration formula (\ref{oneloopfermi}) shows. So, we identify the argument and expand the resulting self-energy amplitude in the decoupling limit to show essentially the same behavior as the suppressed amplitude (\ref{oneloopres}).

\subsubsection{The mass and the momentum dependence of a general amplitude}

The general form of the $n$-loop self-energy amplitude is
\be 
\Sigma_n(p^2) = \int d^4k_1 \dots d^4 k_n P(p,k_1,\dots,k_n) \prod_{j=1}^l \frac{1}{L_j(p,k_1,\dots,k_n)-m_j^2} ,
\ee
where $L_j$ are quadratic functions of the momenta and $m_j$ are the masses of the loop propagating fields. $P$ is a polynomial arising from the spins of the fields. We may absorb the dependence on the couplings to the $P$. This is the general form of any loop order involving fields of any spins.

Using the Feynman parametrization, we can complete the denominator
\be \label{compden}
 \Sigma_n(p^2) = \int d{\bf x} 
  \int d^4k_1 \dots d^4 k_n \frac{\tilde P(k_1,\dots,k_n, p^2) }{[Q(k_1,\dots,k_n) + \Delta_n (p^2)]^l} 
\ee
with the collective notation ${d \bf x}= dx_1 \dots d x_n$ allowing some extra factors as functions of ${\bf x} \equiv \{x_i \}.$ In the integrand, $\tilde P$ is another polynomial and $Q$ is a quadratic function in $k_i$ and $p^2$, whose details are not important here. In this paper, we assume the symmetry $p \leftrightarrow - p$. By completing the square in every $k_i$ and shifting some momenta, we can always single out the dependence in $p^2$ into a unique combination 
\be \label{Delta}
 \Delta_n(p^2) = - f_n({\bf x})p^2 + \sum_i g_{ni} ({\bf x}) m_i^2,
\ee
where $f_n({\bf x}),g_{ni} ({\bf x})$ are the coefficients of $p^2, m_i^2,$ respectively. In the previous example, $f(x)=x(1-x),g(x)=1.$ (Note that once we fix the Feynman parameters, the combination $\Delta_n(p^2)$ is unique. However, there can be other possible parametrizations. Nevertheless, the following argument is valid for any fixed parametrization.)

 We imagine that some of the masses $m_i$ are those of heavy fields in (\ref{UVLag}). It is convenient to introduce the average mass $M$ as 
\be
 g_n({\bf x}) M^2 \equiv \sum_i g_{ni} ({\bf x}) m_i^2,
\ee
with some coefficient $g_n({\bf x})$. If one of the $m_i$ is large, $M$ is also.\footnote{Of course, the coefficients $g_i$ are generic and there should be no miraculous cancellation between any $g_i m_i^2$'s.} 
In the decoupling limit, we thus have $M^2 \gg m_h^2$ and $M^2 \gg p^2$. 
At this moment, the only important thing is the $\Delta_n(p^2)$ in (\ref{Delta}) and its dependence on the large mass $M$,
\be
 \Delta_n(p^2) = - f_n({\bf x})p^2 + g_n({\bf x}) M^2.
\ee 

\subsubsection{The main theorem} 

Note that $\Sigma_n(p^2)$ is of mass dimension two and is a function of $\Delta_n(p^2)$ that is also dimension two. (Explicit factors of $p^2$ in the numerator polynomial $\tilde P$ are traded for $\Delta_n$ through the linear relation $f_n({\bf x}) p^2 = g_n({\bf x}) M^2 - \Delta_n(p^2)$ at fixed Feynman parameters; this is why the coefficients below must be allowed to be as large as ${\cal O}(M^2)$.) After the momentum integration, the amplitude has the form
\be \label{selfeform}
\Sigma_n(p^2)= \int d{\bf x} \left[ A_n ({\bf x})+ c_n ({\bf x}) \Delta_n (p^2) +\left(d_n ({\bf x}) + e_n ({\bf x}) \Delta_n(p^2) \right) \log \frac{B_n({\bf x})}{\Delta_n (p^2)} \right],
\ee
for each $n$, where $c_n ({\bf x}),e_n ({\bf x})$ are order-one constants in the momenta. Strictly speaking, the single-logarithm form (\ref{selfeform}) applies once the subdiagram divergences are subtracted, as described in Section \ref{subsec:subamp} below; before the subtraction, a multi-loop amplitude may contain higher powers of logarithms whose arguments involve the regulator. As we see shortly in the example, the dominant decoupling effect of loop amplitude comes from the renormalization of the main diagram.

From the structure of the integral (\ref{compden}), the argument of the logarithm must be $\Delta_n(p^2)$ as in (\ref{deltadelta}). The coefficient of the logarithm can be a linear polynomial in $(p^2)$, which is expressed as $e_n \Delta_n(p^2)$ and the remainder is absorbed in the dimension two constant $d_n({\bf x})$. The dimensionless constant $c_n ({\bf x})$ can appear, depending on the choice of regularization. In the previous example with cutoff, we had one-loop $n=1$ and  $c_1(x)=-y^2 N_c/(4\pi^2),e_1(x)=3y^2 N_c/(8\pi^2),$ with $B_1=\Lambda^2$ and $d_1=0$. As before, $A_n({\bf x})+ c_n({\bf x}) \Delta_n(p^2)$ is also the most general linear polynomial in $p^2$. A different regularization, such as the dimensional one, would give different values of $c_1(x)$. However, we see that the corresponding term will vanish in the end. The divergence is regularized as $p^2$-independent constants $A_n({\bf x})$ and $B_n({\bf x})$, both of which have mass dimension two.  

The mass correction (\ref{SlidingMass}) that we observe in the experiment is
\be \label{MassCorr} \begin{split}
 \Sigma_n^{\text{ren}}(p^2)&=  \Sigma_n(p^2) - \Sigma_n(m_h^2) - (p^2-m_h^2) \frac{d \Sigma_n}{d p^2}(m_h^2) \\
 &=   \int d{\bf x}   \left[\left(d_n ({\bf x}) + e_n ({\bf x}) \Delta_n(p^2) \right) \log \frac{\Delta_n(m_h^2)}{\Delta_n(p^2)} - f_n ({\bf x})  (p^2-m_h^2)\left(e_n({\bf x}) +\frac{d_n({\bf x})}{\Delta_n(m_h^2)} \right) \right].
 \end{split}
\ee
The scheme-dependent term proportional to $c_n ({\bf x})$ vanishes thanks to the relation (\ref{Deltaprel}), as promised.
The quadratic divergence $A_n$ is canceled in $\Sigma_n(p^2) - \Sigma_n(m_h^2)$ and the divergence in the logarithm $B_n$ is canceled in the first order expansion in $p^2$ \cite{Choi:2023cqs}. 

Consider the case $d_n({\bf x}) = 0$ first.
In the decoupling limit, we have a similar expansion as (\ref{logexp}) 
\be 
\log \Delta_n(p^2) = \log (g_n M^2) -\frac{f_n p^2}{g_n M^2} -\frac{f_n^2 p^4}{2 g_n^2 M^4} +{\cal O}\left( \frac{p^6}{M^6} \right).
\ee
By construction, $g_n M^2 \ne 0.$
Therefore, for each $n$,
\be \label{RPM} \begin{split}
 \Sigma_n^{\text{ren}}(p^2) &=  \int d{\bf x}\, e_n  \bigg[ (g_n M^2-f_n p^2) \left( \frac{f_n (p^2-m_h^2)}{g_n M^2} +  \frac{f_n^2 (p^4 - m_h^4)}{ 2 g_n^2 M^4} + {\cal O}\left( \frac{p^6}{M^6} \right) - {\cal O}\left( \frac{m_h^6}{M^6} \right)   \right)   \\
 & \qquad - f_n (p^2-m_h^2) \bigg] \\
 &=  \int d{\bf x} \frac{e_n f_n^2}{g_n} \left[ - \frac{(p^2-m_h^2)^2}{2 M^2}  +{\cal O}\left( \frac{p^6}{M^4} \right) - {\cal O}\left( \frac{m_h^6}{M^4} \right)   \right] .
\end{split}
\ee
{\em The cancellation of the quadratic terms in $p$ and $m$ takes place,} thanks to the relation (\ref{Deltaprel}),
\be
 \frac{d \Delta_n}{dp^2}(m_h^2) =- f_n({\bf x}), \quad \Delta_n(p^2) - \Delta_n(m_h^2) =- f_n({\bf x})(p^2-m_h^2).
\ee
Naively, the loop correction (\ref{selfeform}) seems proportional to $M^2 \log M^2$.
However, in the physical mass combination (\ref{MassCorr}), the leading order contributions cancel. 

Now consider nonzero $d_n({\bf x})$. The additional contribution to $\Sigma_n^{\text{ren}}(p^2)$ is subdominant 
\be \label{dterm} \begin{split}
\int d{\bf x} & \, d_n \left[  \log \frac{\Delta_n(m_h^2)}{\Delta_n(p^2)} - \frac{f_n(p^2-m_h^2)}{\Delta_n(m_h^2)} \right] \\
 &=  \int d{\bf x}   f_n d_n \left[ \frac{p^2-m_h^2}{g_n M^2}    +\frac{ f_n  (p^4-m_h^4)}{2g_n^2 M^4} + {\cal O}\left( \frac{p^6}{M^6} \right)- {\cal O}\left( \frac{m_h^6}{M^6} \right) - \frac{p^2-m_h^2}{g_n M^2-f_n m_h^2} 
  \right] \\
 & = \int d{\bf x} \frac{  f_n^2  }{g_n^2}d_n \left[ \frac{(p^2-m_h^2)^2}{2M^4}+ {\cal O}\left( \frac{p^6}{M^6} \right)- {\cal O}\left( \frac{m_h^6}{M^6} \right) \right].
\end{split}
\ee
In each line, we expanded the logarithm and the denominator in the decoupling limit.
So, at most this contribution (\ref{dterm}) is the same order as that without it (\ref{RPM}), if $d_n$ is ${\cal O}(M^2)$.

Therefore, the total amplitude is suppressed as 
\be 
 \Sigma_n^{\text{ren}}(p^2) = {\cal O} \left( \frac{(p^2-m_h^2)^2}{M^2} \right).
\ee
In the heavy mass limit, the corresponding mass correction vanishes
\be
 M \to \infty \quad \Longrightarrow \quad  \Sigma_n^{\text{ren}} (p^2) \to 0.
\ee
This is the suppression expected from the Appelquist--Carazzone decoupling theorem, which we make explicit for the scalar mass, the relevant parameter that the original statement sets aside.
We expect a similar behavior for other super-renormalizable operators, such as a cubic scalar coupling, although the dimensional counting used above does not directly apply when a dimensionful coupling is present. It exhibits power-law running, as seen from the dependence on $p^2$ \cite{Choi:2023mma}. The dependence on the combination $p^2-m_h^2$ is expected because the reference point is $p^2=m_h^2$. The overall sign is determined by $-\int d{\bf x} e_n({\bf x}) f_n({\bf x})^2/g({\bf x})$.
Light fields can correct the mass sizably, but it is not hierarchical.\footnote{There can be a possible mixing of the light scalar $\phi$ and the heavy scalar $X$ through the loop correction. The corresponding amplitude has a 1PI diagram with one external $\phi$ and one $X$. Ref. \cite{Kazama:1981fx} has shown that, after the diagonalization, the mixing is suppressed as ${\cal O}(m_h^2/M_s^2)$.}

Three remarks are in order regarding the status of this result. First, as stated above, the limit is taken with the couplings held fixed. Second, the higher powers of logarithms that appear at higher loop orders weaken the suppression at most by powers of $\log M^2$. This follows from the integral form of the Taylor remainder,
\be \label{remainderint}
 \Sigma_n^{\text{ren}}(p^2) = \int_{m_h^2}^{p^2} ds\, (p^2 - s)\, \frac{d^2 \Sigma_n}{ds^2}(s),
\ee
where the second derivative of the subdiagram-renormalized amplitude has mass dimension $-2$ and overall degree of divergence $-2$, hence is ultraviolet convergent by power counting. The argument applies to any term of the form $\Delta_n^a \log^b \Delta_n$ and not only to the single-logarithm structure (\ref{selfeform}). By the large-mass expansion, the second derivative separates into a part in which the heavy lines are hard, which carries an explicit $1/M^2$ up to powers of $\log M^2$, and a part built from the light lines alone, which carries no dependence on $M$ at all. Only the first part is the subject of the decoupling statement. The second is the ordinary light-field correction discussed below, which is not suppressed but is not enhanced by the heavy scale either, and which therefore does not reintroduce a hierarchy.

The conclusion is supported independently by the asymptotic large-mass expansion of Feynman integrals \cite{Gorishnii:1989dd, Smirnov:1990rz, Smirnov:2002pj}: the hard regions of the loop momenta produce a polynomial in $p^2$ whose coefficients contain $M^2$ and $\log M^2$ and are removed exactly by the double subtraction in (\ref{MassCorr}), while the soft regions carry explicit powers of $1/M^2$ from the Taylor-expanded heavy propagators. Rigorous decoupling theorems for BPHZ-renormalized amplitudes, with the subtractions performed at the origin, have been established in \cite{Ambjorn:1979nv, Manoukian:1981pb, Manoukian:1983book}.\footnote{When massless fields circulate in the loop, the momentum subtraction requires the Lowenstein--Zimmermann modification \cite{Lowenstein:1975ps, Lowenstein:1975rg}. This does not affect the heavy-field suppression, since the thresholds involving the heavy field open far above $p^2 = m_h^2$.} The contribution of the present formulation is to make the decoupling manifest in the on-shell scheme at the physical reference point, where the subtraction terms are the complete counterterms and no separate matching is involved.

It is useful to locate the present statement against this earlier literature. Appelquist and Carazzone \cite{Appelquist:1974tg} argued that the effect of a heavy field on light-particle Green functions is suppressed by inverse powers of $M$, provided that the finite parts of the divergent integrals are defined by subtractions at the light scale and that no vertex grows as $M \to \infty$. Their theorem is stated up to the renormalization of the couplings, the masses and the field normalizations of the light theory, and the scalar mass is exactly what that proviso sets aside. Their counting also rests on the identity that the dimensionality of a graph equals its degree of divergence, which holds only when every coupling is dimensionless. In the form in which it is stated, therefore, the theorem does not speak about the correction to $m^2$.

Schimert and Chiang \cite{Schimert:1983fb} made the argument rigorous diagram by diagram, building on the recursive construction of Caswell and Kennedy \cite{Caswell:1981ek}. Replacing minimal subtraction by a Taylor momentum subtraction performed at a reference point of the order of the light-mass scale, an operation they show to be equivalent to adding finite counterterms to the Lagrangian, they proved by induction on the number of loops that the heavy-mass dependence of an arbitrary diagram is bounded by a negative power of $M$, up to one factor of $\log M$ per loop. Their prescription assigns the exponent $-1$ to each divergent subdiagram that the subtraction has made finite, for the same dimensional reason recalled above, and the decoupling theorem follows as a corollary. Their subtraction operator is the one appearing in (\ref{SlidingMass}), and their bound formally covers the two-point function as the case of two external lines. 
The behavior established below is thus consistent with, and largely implied by, their power counting.
Two things are left open for our purpose. The subtracted quantities $\Sigma_n(m_h^2)$ and $d\Sigma_n/d(p^2)(m_h^2)$ enter their construction as inputs fixed by hand, so the theorem says nothing about whether what survives the subtraction is the whole of the physical mass correction or only a fragment of it. And the exponent $-1$ is not saturated here, since the $p \leftrightarrow -p$ symmetry assumed below (\ref{compden}) forbids the odd power of momentum that the naive counting allows. Their analysis is also restricted to Euclidean momenta, whereas the reference point $p^2 = m_h^2$ used here is timelike.

The present formulation supplies both missing pieces. Identifying the observable with $m^2(p^2)$ of (\ref{totmassub}), which is what the $S$ matrix measures, the whole of the loop correction to the scalar mass is the renormalized self-energy (\ref{MassCorr}), and it is suppressed as $(p^2-m_h^2)^2/M^2$ at every order in the couplings. The quantity that the earlier treatments left as an input is not separately observable, and the quantity that they bounded is not a fragment but the entire correction. What the estimate (\ref{UVsensitivity}) records is then an artifact of a scale-independent subtraction rather than a sensitivity of the low-energy theory to the ultraviolet.

\subsubsection{Subamplitude renormalization} \label{subsec:subamp}

It remains to show that the subdiagram divergence does not affect this result. We use the subdiagram renormalization seen in Section \ref{secEff}, which is essentially equivalent to the BPHZ renormalization \cite{Bogoliubov:1957gp, Hepp:1966eg, Zimmermann:1967, Zimmermann:1969jj, Lowenstein:1975ps, Lowenstein:1975rg}. If we have $l$-loop divergence in the subdiagram $\gamma$, we separate the corresponding loop with variables $k_1,\dots k_l$, without loss of generality
\be \label{mainamp}
\Sigma_{n-l}(p^2) =   \int d^4k_{l+1} \dots d^4 k_n A(p,k_{l+1},\dots,k_n) P(p,k_{l+1},\dots,k_n) \prod_{j} \frac{1}{L_j(p,k_{l+1},\dots,k_n)-m_j^2} ,
\ee
with
\be \label{subamp}
A(p,k_{l+1},\dots,k_n)  = \int d^4k_1 \dots d^4 k_l P(p,k_1,\dots,k_n) \prod_{j} \frac{1}{L_j(p,k_1,\dots,k_n)-m_j^2}.
\ee
This is always possible in the renormalizable theory, and we consider the lowest-loop-order divergent subamplitude \cite{Choi:2025tus}.
The important point is that the main amplitude (\ref{mainamp}) does not depend on the subamplitude momenta $k_1, \dots,k_l$. As in renormalizing the main amplitude, we put the Taylor expansion operator acting on the amplitude $A(p,k_{l+1},\dots,k_n)$,
\be \label{subrenamp} \begin{split}
 \Sigma_{n-l}^{\text{sub}} (p^2) &= (1-t^D)  A(p,k_{l+1},\dots,k_n)\\
 &=  \int d^4k_{l+1} \dots d^4 k_n (1-t^D)   P(p,k_{l+1},\dots,k_n) \\ &\quad \times \prod_{j} \frac{1}{L_j(p,k_{l+1},\dots,k_n)-m_j^2}.
\end{split}
\ee
Now, the renormalized amplitude (\ref{subrenamp}) is the sum of formally $(n-l)$-loop order amplitudes having the same form as the original one (\ref{SMQ}). By induction, this amplitude also exhibits decoupling behavior if the superheavy field is inside it, as shown in (\ref{subrenamp}). If superheavy fields are inside the subamplitude (\ref{subamp}), we apply the conventional Appelquist--Carazzone decoupling theorem. Explicitly, the subtracted subamplitude $(1-t^D)A$ vanishes to the required order at the reference point and re-enters (\ref{mainamp}) as a convergent insertion of formally lower loop order, carrying the same dependence on $\Delta$. The induction hypothesis then applies to the resulting $(n-l)$-loop object, so the heavy-mass suppression $(p^2-m_h^2)^2/M^2$ is preserved at each step.

%%%%%%%%%%%%%%%%%%%%%%%%%%%%%%%
\subsection{Decoupling of a scalar at two loops}

\begin{figure}[t]
\begin{center}
\includegraphics[scale=0.75]{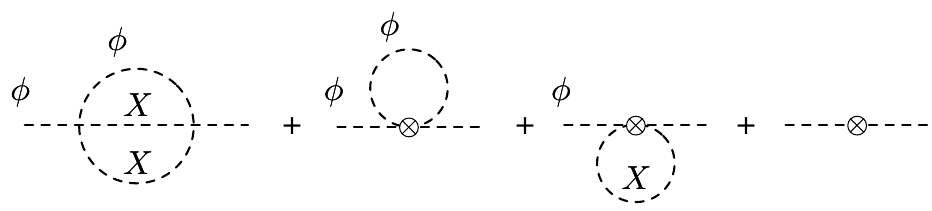}
\end{center}
\caption{The two-loop self-energy of a scalar $\phi$ involving a UV scalar $X$. As in the previous case of the fermion loop, the renormalized combination vanishes in the decoupling limit. Interestingly, the one-loop counterterms vanish due to the same structure as in (\ref{Sigma1}).  \label{fig:phitwoloop} }
\end{figure}

Another example is a two-loop contribution including the heavy scalar $X$, which is drawn in Fig. \ref{fig:phitwoloop}. We will see that the decoupling of heavy fields is primarily driven by the main divergence, rather than the subdiagram ones.
%\footnote{A one-loop contribution trivially vanishes because the self-energy $\Sigma_\lambda(p^2)$ is independent of $p$ thus the combination (\ref{SlidingMass}) is zero \cite{Peskin:1995ev}.} 
Its amplitude is 
\be
 -i\Sigma_2(p^2) = \frac{(-i \kappa)^2 }{2}\int \frac{d^4 k}{(2\pi)^4}  \int \frac{d^4 l}{(2\pi)^4}\frac{i}{k^2-m_h^2}\frac{i}{l^2-M_s^2}  \frac{i}{(p+k+l)^2-M_s^2}  ,
\ee
where the subscript in $\Sigma_2(p^2)$ denotes the loop order through the coupling $\kappa$ between $\phi$ and $X$. 

This amplitude has subamplitude divergence, which is not distinguished from the main divergence. This is partly due to the overlapping loop structure and is closely related to the fact that it has a vanishing one-loop. The alleged places are the large momentum regime of each subamplitude that behaves as 
\be 
\quad k_E\sim \Lambda,\ l_E\sim \Lambda \text{ or } \ k_E+l_E \sim \Lambda: \quad \Sigma_2(p^2) \sim \Lambda^2 \log \Lambda,
\ee 
in each case.

For instance, for the subdiagram divergence as $l_E \sim \Lambda,$ we renormalize the corresponding subamplitude as in (\ref{OneComp}),
\be 
\begin{split}
 \Sigma^{\text{sub}}_2(p^2) &\equiv -\frac{\kappa^2 }{2}  \int \frac{d^4 k}{(2\pi)^4} D(k^2) \\
 & \qquad \times  \int \frac{d^4 l}{(2\pi)^4}  D
 _{M_s}(l^2)  \Big[  D_{M_s}((p+k+l)^2) - D_{M_s}((p+k+l)^2) _{(p+k)^2=\mu_{\phi\phi}^2} \Big] \\
 &=  -\frac{\kappa^2 }{2} \int \frac{d^4 k}{(2\pi)^4}  \frac{1}{k^2-m_h^2}\Big[ V(p+k) - V(p+k)_{(p+k)^2=\mu_{\phi\phi}^2} \Big]  \\
 &=  -\frac{\kappa^2 }{2} \int \frac{d^4 k}{(2\pi)^4}  \frac{1}{k^2-m_h^2} \int_0^1 dx \log \frac{M_s^2 - x(1-x)(p+k)^2}{M_s^2-x(1-x) \mu_{\phi\phi}^2}.
\end{split}
\ee
The logarithmic divergence is canceled by the same quantity at the reference point $(p+k)^2=\mu_{\phi\phi}^2$.

However, the subtracted term is constant, because the rest part $D(k^2)$ does not depend on the external momentum, so the overall renormalization removes it
\be \label{overallm2}
\begin{split}
 \Sigma_2^{\text{ren}} (p^2)   &=  \Sigma_2^{\rm sub} (p^2) - \Sigma_2^{\rm sub} (m_h^2) - (p^2-m_h^2) \frac{d \Sigma_2^{\rm sub}}{dp^2} (m_h^2) \\
 & = -\frac{\kappa^2 }{2} \int \frac{d^4 l}{(2\pi)^4}  \int \frac{d^4 k}{(2\pi)^4} \frac{1}{l^2-M_s^2}  \frac{1}{k^2-m_h^2} \bigg[   \frac{1}{(p+k+l)^2-M_s^2} \\
 &\quad \qquad - \frac{1}{(p+k+l)^2-M_s^2}\bigg|_{p^2=m_h^2}-(p^2-m_h^2)  \frac{1}{\big((p+k+l)^2-M_s^2\big)^2}\bigg|_{p^2=m_h^2} \bigg] .
\end{split}
\ee
The second and the third terms are the ``counterterms'' for the main divergence, which is collectively depicted as the last diagram in Fig. \ref{fig:phitwoloop}. Note that the second term removes the subdiagram divergences of the $k$ and $l$ loops at the same time. To see this, we may perform part of the integral
\begin{align}
 \Sigma_2^{\text{ren}} (p^2)  
 & = -\frac{\kappa^2 }{2} \int \frac{d^4 l}{(2\pi)^4} \frac{1}{l^2-M_s^2} \Big[ V(p+l) - V(p+l)_{p^2=m_h^2} \Big] + {\cal O} (p^2-m_h^2) \\
 & = -\frac{\kappa^2 }{2} \int \frac{d^4 k}{(2\pi)^4} \frac{1}{k^2-m_h^2} \Big[ V(p+k) - V(p+k)_{p^2=m_h^2} \Big] + {\cal O} (p^2-m_h^2) \\
& = -\frac{\kappa^2 }{2} \int \frac{d^4 k}{(2\pi)^4} \frac{1}{k^2-m_h^2} \Big[ V(p) - V(p)_{p^2=m_h^2} \Big] + {\cal O} (p^2-m_h^2).
\end{align}
In the last line, the dummy integral variable inside $V(p)$ is $k+l$ with one, say, $k$ fixed. In each line, the symbol $V$ collectively denotes the one-loop bubble formed by the two lines carrying the loop momentum that becomes large, with the corresponding internal masses; the subtracted constant is common to each region because the logarithmically divergent part of the bubble is independent of the masses and of the external argument. The terms denoted ${\cal O} (p^2-m_h^2)$ are finite and are restored in the complete evaluation (\ref{indiSigma}) below.
Using (\ref{overallm2}), there are no subamplitude divergences in either the large-$l_E$ or large-$k_E$ regime.

The overall amplitude is evaluated as follows.
Introducing the Feynman parameters, the integrand becomes
\be 
 \int_0^1 dx \int_0^x dy \frac{-i}{[x (l^2-M_s^2)+y((p+k+l)^2-M_s^2)+(1-x-y)(k^2-m_h^2)]^3}.
\ee
The denominator becomes a quadratic form in $k,l$ and $p$. Further completing the squares and shifting the momenta $k,l,p$ makes the integrand of the form (\ref{compden}), 
\be
 \Sigma_2(p^2) = -  \frac{\kappa^2 }{2}\int \frac{d^4 k}{(2\pi)^4}  \int \frac{d^4 l}{(2\pi)^4} \int_0^1 dx \int_0^x dy \frac{1}{(a k^2 + b l^2 + f p^2 - g_1 M_s^2 -g_2 m_h^2)^3}.
\ee
determining 
\be \Delta_n(p^2) = -fp^2 + g_1 M_s^2 + g_2 m_h^2,
\ee 
with $a,b,f,g_1,g_2$ being functions of $x,y$. This $\Delta_n(p^2)$ is dominated by $M_s^2 \gg m_h^2$, but the $p^2$-dependence is important. After performing the Wick rotations for both variables, the integrand schematically has the form
\be \label{Euc2l}
\begin{split}
 \int d^4 k_E &\int   d^4 l_E \frac{1}{(k_E^2+l_E^2+\Delta_n(p^2))^3} \\
 & =\pi^4 \int_0^{\Lambda_1^2}  d (k_E^2)k_E^2 \int_0^{\Lambda_2^2}   d (l_E^2)l_E^2 \frac{1}{(k_E^2+l_E^2+\Delta_n(p^2))^3}  \\
 & =\pi^4 \int_0^{\Lambda_1^2} d (k_E^2)k_E^2  \frac{\Lambda_2^4}{2( k_E^2+\Delta_n(p^2))(k_E^2+\Delta_n(p^2)+\Lambda_2^2)^2} \\
 & =\pi^4 \int_0^{\Lambda_1^2}  d (k_E^2)k_E^2 \frac{1}{2( k_E^2+\Delta_n(p^2))} \\
 & = \frac{\pi^4}{2} \left(  \Lambda_1^2 - \Delta_n(p^2) \log \frac{\Lambda_1^2}{\Delta_n(p^2)} \right).
\end{split}
\ee
The result has exactly the same form as (\ref{selfeform}) with $c_n(x,y)=d_n(x,y)=0$. Here, we have introduced the cutoffs $k_E^2=\Lambda_1^2$ and $l_E^2=\Lambda_2^2$, resulting in no linear term in $\Delta_n$. Other regularizations may produce a linear $c_n (x,y) \Delta_n$ term. Also, the procedure (\ref{Euc2l}) shows that the self-energy depends on the order of the integrations; however, as we see below, such dependence disappears, and we have the unique result as the physical mass correction. It is straightforward to generalize this result to higher-order loops.

Therefore,
\be \label{indiSigma}
 \Sigma_2^{\text{ren}}(p^2) \propto \int_0^1 dx \int_0^x dy  \left [ \Delta(p^2) \log \frac{\Delta(m_h^2)}{\Delta(p^2)} - f \cdot (p^2-m_h^2) \right] .
\ee
This is of the form (\ref{MassCorr}). It follows that the heavy scalar $X$ decouples in the limit $M_s \to \infty$.

%%%%%%%%%%%%%%%%%%%%%%%%%%%%%%%
\subsection{Decoupling with intermediate mass scales} \label{subsec:intermscale}

The general formula (\ref{Delta}) introduces the average heavy mass $M^2$ and proceeds in the limit $M^2 \gg m_h^2, p^2$. One may ask whether this simplification remains valid when the loop contains widely separated mass scales, in particular when an intermediate field of mass $m_Y$ satisfying $m_h \ll m_Y \ll M$ mediates the coupling of the heavy field $X$ to the external scalar. This is the relevant scenario for any UV completion in which $X$ does not couple directly to the Higgs but interacts only through an intermediate field, generating chain contributions of the type identified as a residual source of $M^2$ sensitivity at higher loops.

Consider the representative chain in which the heavy field of mass $M$ runs in a sub-loop and the intermediate field of mass $m_Y$ carries the external momentum dependence. Combining the two propagators with Feynman parameter $x$ weighting the heavy line, the relevant combination reads
\be \label{eq:chainDelta}
\Delta(p^2) = x M^2 + (1-x) m_Y^2 - f(x) p^2 ,
\ee
which is a particular realization of the general form (\ref{Delta}) with two distinct mass scales appearing in $\sum_i g_{ni} m_i^2$. The Feynman parameter $x$ partitions the integration into two qualitatively different regions and the partition is structural rather than accidental.

In the bulk region $x \gtrsim m_Y^2/M^2$, the heavy mass dominates the local integrand $\Delta \approx x M^2$. The integrand exhibits the expected $M^2$ growth and the Taylor subtraction at $p^2 = m_h^2$ removes the constant and linear-in-$p^2$ pieces by the same mechanism that produced (\ref{RPM}). The surviving Taylor remainder is parametrically $(p^2 - m_h^2)^2/(xM^2)$, and integrating $1/x$ over the bulk from $m_Y^2/M^2$ to $1$ gives a contribution of order $(p^2 - m_h^2)^2/M^2$ multiplied by a logarithm $\log(M^2/m_Y^2)$.

In the small-$x$ corner $x \lesssim m_Y^2/M^2$, the situation is qualitatively different. The heavy mass is suppressed by the small Feynman weight, $x M^2 \lesssim m_Y^2$, and the local denominator becomes $\Delta \approx m_Y^2$. The integrand is locally enhanced by $1/m_Y^2$ rather than $1/M^2$. However the width of the corner is itself $\Delta x \sim m_Y^2/M^2$, and the product
\be
 \Delta x \cdot \frac{1}{m_Y^2} = \frac{m_Y^2}{M^2} \cdot \frac{1}{m_Y^2} = \frac{1}{M^2}
\ee
reproduces the same $1/M^2$ scaling as the bulk. The corner cannot enhance the result over the heavy-mass region because the width of the region is itself set by the same mass ratio.

The structural content of this observation is that the regions of Feynman parameter space in which the heavy mass $M$ enters $\Delta(p^2)$ as a large scale are exactly the regions in which the BPHZ subtraction removes the $M^2$ growth, while the regions in which $\Delta$ is dominated by light masses are exactly the regions in which there is no $M^2$ piece to remove or enhance. The combination of a heavy-mass source with a small-$\Delta$ denominator, which would generate dangerous $M^2/m_Y^2$ chain enhancements, simply does not occur in the renormalized integrand. The Feynman parametrization distributes the loop contributions so that ``heavy mass present in $\Delta$'' and ``small-$\Delta$ enhancement'' are mutually exclusive conditions.

This argument extends inductively to arbitrarily long chains of intermediate scales $m_h \ll m_{Y_1} \ll \cdots \ll m_{Y_k} \ll M$. Each Feynman parameter associated with a heavy line carries the same partition structure: bulk regions where the heavy line participates in $\Delta$ have $M^2$ growth removed by the Taylor subtraction, corner regions where it does not participate have no $M^2$ piece to begin with. The result remains
\be \label{eq:chaindecoup}
 \Sigma_n^{\text{ren}}(p^2) = {\cal O} \left( \frac{(p^2 - m_h^2)^2}{M^2} \right) ,
\ee
up to logarithms $\log(M^2/m_{Y_i}^2)$ that do not alter the parametric scaling.

To make the partition explicit, evaluate the representative renormalized
contribution. With the affine combination (\ref{eq:chainDelta}) and the
$e_n \Delta \log \Delta$ structure of (\ref{selfeform}), the double subtraction
gives, for each Feynman parameter,
\be
\begin{split}
 (1-t^2_{m_h})\big[\Delta \log \Delta\big]
  & = \Delta \log \frac{\Delta}{\Delta_0} - (\Delta - \Delta_0) \\
  & = \frac{f(x)^2 (p^2-m_h^2)^2}{2\,\Delta_0} + {\cal O}\!\left(\frac{(p^2-m_h^2)^3}{\Delta_0^2}\right),
\end{split}
\ee
with $\Delta_0 = \Delta(m_h^2)$. Integrating over the Feynman parameter with
$f(x) = x(1-x)$,
\be
\begin{split}
 \Sigma^{\text{ren}}(p^2) & \supset
   \frac{(p^2-m_h^2)^2}{2} \int_0^1 dx\,
   \frac{x^2(1-x)^2}{x M^2 + (1-x) m_Y^2} \\
   & = \frac{(p^2-m_h^2)^2}{24 M^2}
     \left[ 1 - \frac{3 m_Y^2}{M^2} + {\cal O}\!\left(\frac{m_Y^4}{M^4}\log \frac{M^2}{m_Y^2}\right) \right].
\end{split}
\ee
The result is ${\cal O}((p^2-m_h^2)^2/M^2)$, the intermediate scale entering
only through corrections suppressed by $m_Y^2/M^2$. The integral is dominated by
the bulk $x \sim {\cal O}(1)$, where the denominator is $x M^2$; the small-$x$
corner is doubly suppressed because the numerator $f(x)^2 = x^2(1-x)^2$ vanishes
there, so it contributes only at ${\cal O}(m_Y^4/M^6)$. No $1/m_Y^2$ appears.

For the genuine chain, in which $X$ enters only through a sub-loop on the
intermediate line, the mechanism is sharper. Applying the on-shell subtraction
(\ref{SlidingMass}) to the $Y$ two-point function, the heavy $X$ insertion is
renormalized as
\be
 \Pi_X^{\text{ren}}(q^2) = (1-t^2_{m_Y}) \Pi_X(q^2)
   = {\cal O}\!\left(\frac{(q^2-m_Y^2)^2}{M^2}\right),
\ee
which is the present theorem applied to $Y$ in place of $\phi$. At fixed order
the chain contributes to the $\phi$ self-energy through the insertion
\be
 \frac{1}{q^2-m_Y^2}\, \Pi_X^{\text{ren}}(q^2)\, \frac{1}{q^2-m_Y^2}
   = {\cal O}\!\left(\frac{1}{M^2}\right),
\ee
where the double zero of $\Pi_X^{\text{ren}}$ at $q^2=m_Y^2$ cancels the double
pole of the intermediate line. The would-be $M^2/m_Y^2$ enhancement requires a
heavy-mass numerator sitting on a near-on-shell light denominator, and this
cannot occur, because the heavy insertion vanishes quadratically exactly where
the light line goes on-shell. The remaining $\phi$-loop integral is then
${\cal O}(1/M^2)$, and the external subtraction (\ref{MassCorr}) renders it
${\cal O}((p^2-m_h^2)^2/M^2)$. The argument iterates along a chain
$m_h \ll m_{Y_1} \ll \cdots \ll M$: each heavy insertion carries two powers of
the adjacent off-shellness and defuses the pole of the line it sits on.

The form $\Pi_X^{\text{ren}}(q^2) = {\cal O}((q^2-m_Y^2)^2/M^2)$ is an expansion valid for $q^2 \ll M^2$, while the momentum $q$ of the intermediate line is integrated over its whole range. The two regions must therefore be treated separately, and both give the same scaling. For $q^2 \lesssim m_Y^2$ the expansion applies, the double zero of $\Pi_X^{\text{ren}}$ at $q^2 = m_Y^2$ cancels the double pole of the intermediate line, and the insertion is ${\cal O}(1/M^2)$ as stated. For $q^2 \sim M^2$ the expansion is not valid and the renormalized insertion grows, behaving as $\Pi_X^{\text{ren}}(q^2) \sim q^2$ up to logarithms, which is the ordinary large-momentum behavior of a renormalized two-point function. But in that region the intermediate propagators are themselves of order $1/q^2$, so that
\be
 \frac{1}{q^2-m_Y^2}\, \Pi_X^{\text{ren}}(q^2)\, \frac{1}{q^2-m_Y^2}
  \sim \frac{q^2}{q^4} \sim \frac{1}{M^2}
\ee
up to logarithms, and the contribution is again ${\cal O}(1/M^2)$. The intermediate scale $m_Y$ is negligible throughout this region and cannot appear in a denominator. The dangerous combination $M^2/m_Y^2$ would require the heavy insertion to be large while the intermediate line is close to its pole, and the two regions just described are precisely the statement that this cannot happen. In the soft region the insertion vanishes quadratically where the line goes on shell, and in the hard region the line is far off shell. The remaining loop integral is then of order $1/M^2$ in either case, and the external subtraction (\ref{MassCorr}) renders it ${\cal O}((p^2-m_h^2)^2/M^2)$.

Therefore, the average-mass formulation introduced in (\ref{Delta}) is parametrically justified and does not require the loop content to involve only one mass scale. The footnote disclaimer following (\ref{Delta}) about ``no miraculous cancellation between any $g_i m_i^2$'s'' is sharpened by this analysis: the absence of cancellation between heavy and light contributions is structural, because the Feynman parameter weighting prevents the two from coexisting in the same region of integration.

Chain diagrams with intermediate fields do not regenerate the $M^2$ sensitivity that would otherwise survive the on-shell BPHZ renormalization. This is in contrast to the $\overline{\rm MS}$ matching procedure, in which each intermediate threshold produces a matching coefficient containing an $M^2 \log(M^2/m_Y^2)$ piece that is absorbed into a redefinition of the effective theory parameter at that threshold. The matching coefficient is computable in any specific UV completion but its size scales with the unknown UV coupling structure, so the absorbed piece is parametrically unbounded without further input. The on-shell BPHZ subtraction at the physical pole acts directly on the Feynman-parameterized integrand and removes the same piece without intermediate matching, in a manner that does not depend on the detailed coupling structure of the UV completion. The decoupling (\ref{eq:chaindecoup}) is a universal parametric statement about how the heavy mass enters the renormalized observable, valid regardless of whether the coupling to the Higgs is direct or mediated by intermediate fields.

In summary, the scalar mass is not sensitive to UV physics because the correction of the heavy field is suppressed, and this suppression survives in the presence of intermediate mass scales. The renormalized physical quantity is regularization-independent and the correct identification of physical observables is essential.

%%%%%%%%%%%%%%%%%%%%%%%%%%%%%%%%%%%%
\section{Conclusions}

We have revisited the technical hierarchy problem by distinguishing what is observable from what is a parametrization. The bare mass and the loop corrections are never separately observable; only the combination (\ref{totmassub}) is. Defining the theory by the pole mass $m_h^2$ through the relation (\ref{baremassagain}), the renormalized loop corrections are finite, independent of the regularization and admit a well-defined perturbative expansion. No fine-tuning between independently known quantities is involved, because the apparent cancellation reflects a change of basis among the parameters of the same theory.

The remaining question is the size of the renormalized corrections themselves. We have shown, for the general loop correction involving fields of arbitrary spins and with the couplings held fixed in the heavy-mass limit, that the renormalized self-energy involving a heavy field of mass $M$ is suppressed as $(p^2-m_h^2)^2/M^2$, making the Appelquist--Carazzone
decoupling theorem explicit for the scalar mass, the relevant parameter that its original statement sets aside. The suppression survives in the presence of intermediate mass scales, where chain contributions might be expected to regenerate the $M^2$ sensitivity. Therefore, once the tree-level mass is made small, it remains small against loop corrections, and the scalar mass is insensitive to UV physics, as required for a well-defined effective field theory.
The assumption that the couplings are held fixed is not innocuous. An ultraviolet completion in which the light mass is itself determined by heavy parameters, as in doublet-triplet splitting, violates it, and the question of the origin of the tree-level mass lives there. The origin of the smallness of the tree-level mass remains an open question that our argument does not address.

\subsection*{Acknowledgments}
The author thanks Jong-Hyun Baek, Sungwoo Hong, Hyung-Do Kim, Bumseok Kyae, Hyeseon Im, Stefan Groot Nibbelink, Hans-Peter Nilles, Ruiwen Ouyang, Mu-In Park, Jaewon Song and Piljin Yi for discussions. 

\section*{Declarations}

\subsection*{Funding}
This work is partly supported by the grant RS-2023-00277184 of the National Research Foundation of Korea.

\subsection*{Availability of data and material}
Data sharing is not applicable to this article, as no datasets were generated or analyzed. All results are derived analytically and are
contained within the article. 

\subsection*{Competing interests}
The author declares no competing interests.


\begin{thebibliography}{99}

%\cite{Choi:2023cqs}
\bibitem{Choi:2023cqs}
K.-S.~Choi,
``On the observables of renormalizable interactions,''
J. Korean Phys. Soc. \textbf{84} (2024) no.8, 591-595
doi:10.1007/s40042-024-01025-7
[arXiv:2310.00586 [hep-ph]].

%\cite{Choi:2023mma}
\bibitem{Choi:2023mma}
K.-S.~Choi,
``Exact renormalization of the Higgs field,''
Phys. Rev. D \textbf{109} (2024) no.7, 076008
doi:10.1103/PhysRevD.109.076008
[arXiv:2310.10004 [hep-th]].

%\cite{Georgi:1974yw}
\bibitem{Georgi:1974yw}
H.~Georgi and A.~Pais,
``Calculability and Naturalness in Gauge Theories,''
Phys. Rev. D \textbf{10} (1974), 539
doi:10.1103/PhysRevD.10.539

%\cite{Gildener:1976ai}
\bibitem{Gildener:1976ai}
E.~Gildener,
``Gauge Symmetry Hierarchies,''
Phys. Rev. D \textbf{14} (1976), 1667
doi:10.1103/PhysRevD.14.1667

%\cite{Gildener:1979dd}
\bibitem{Gildener:1979dd}
E.~Gildener,
``Gauge Symmetry Hierarchies Revisited,''
Phys. Lett. B \textbf{92} (1980), 111-114
doi:10.1016/0370-2693(80)90316-0

%\cite{Weinberg:1978ym}
\bibitem{Weinberg:1978ym}
S.~Weinberg,
``Gauge Hierarchies,''
Phys. Lett. B \textbf{82} (1979), 387-391
doi:10.1016/0370-2693(79)90248-X

%\cite{Veltman:1980mj}
\bibitem{Veltman:1980mj}
M.~J.~G.~Veltman,
``The Infrared - Ultraviolet Connection,''
Acta Phys. Polon. B \textbf{12} (1981), 437
Print-80-0851 (MICHIGAN).

%\cite{Natale:1982mt}
\bibitem{Natale:1982mt}
A.~A.~Natale and R.~C.~Shellard,
``The Gauge Hierarchy Problem,''
J. Phys. G \textbf{8} (1982), 635
doi:10.1088/0305-4616/8/5/005

%\cite{Susskind:1982mw}
\bibitem{Susskind:1982mw}
L.~Susskind,
``The Gauge Hierarchy Problem, Technicolor, Supersymmetry, and All That,''
Phys. Rept. \textbf{104} (1984), 181-193
doi:10.1016/0370-1573(84)90208-4

%\cite{Wetterich:1983bi}
\bibitem{Wetterich:1983bi}
C.~Wetterich,
%``Fine Tuning Problem and the Renormalization Group,''
Phys. Lett. B \textbf{140} (1984), 215-222
doi:10.1016/0370-2693(84)90923-7
%75 citations counted in INSPIRE as of 25 Mar 2025

%\cite{Hamada:2012bp}
\bibitem{Hamada:2012bp}
Y.~Hamada, H.~Kawai and K.~y.~Oda,
``Bare Higgs mass at Planck scale,''
Phys. Rev. D \textbf{87} (2013) no.5, 053009
[erratum: Phys. Rev. D \textbf{89} (2014) no.5, 059901]
doi:10.1103/PhysRevD.87.053009
[arXiv:1210.2538 [hep-ph]].

%\cite{Feng:2013pwa}
\bibitem{Feng:2013pwa}
J.~L.~Feng,
``Naturalness and the Status of Supersymmetry,''
Ann. Rev. Nucl. Part. Sci. \textbf{63} (2013), 351-382
doi:10.1146/annurev-nucl-102010-130447
[arXiv:1302.6587 [hep-ph]].

%\cite{Farina:2013mla}
\bibitem{Farina:2013mla}
M.~Farina, D.~Pappadopulo and A.~Strumia,
``A modified naturalness principle and its experimental tests,''
JHEP \textbf{08} (2013), 022
doi:10.1007/JHEP08(2013)022
[arXiv:1303.7244 [hep-ph]].

%\cite{Wells:2013tta}
\bibitem{Wells:2013tta}
J.~D.~Wells,
``The Utility of Naturalness, and how its Application to Quantum Electrodynamics envisages the Standard Model and Higgs Boson,''
Stud. Hist. Phil. Sci. B \textbf{49} (2015), 102-108
doi:10.1016/j.shpsb.2015.01.002
[arXiv:1305.3434 [hep-ph]].

%\cite{Hebecker:2020aqr}
\bibitem{Hebecker:2020aqr}
A.~Hebecker,
``Lectures on Naturalness, String Landscape and Multiverse,''
[arXiv:2008.10625 [hep-th]].


%\cite{Mooij:2021ojy}
\bibitem{Mooij:2021ojy}
S.~Mooij and M.~Shaposhnikov,
``QFT without infinities and hierarchy problem,''
Nucl. Phys. B \textbf{990} (2023), 116172
doi:10.1016/j.nuclphysb.2023.116172
[arXiv:2110.05175 [hep-th]].

%\cite{Aoki:2012xs}
\bibitem{Aoki:2012xs}
H.~Aoki and S.~Iso,
``Revisiting the Naturalness Problem -- Who is afraid of quadratic divergences? --,''
Phys. Rev. D \textbf{86} (2012), 013001
doi:10.1103/PhysRevD.86.013001
[arXiv:1201.0857 [hep-ph]].

%\cite{Jegerlehner:2013cta}
\bibitem{Jegerlehner:2013cta}
F.~Jegerlehner,
``About the role of the Higgs boson in the evolution of the early universe,''
Acta Phys. Polon. B \textbf{45} (2014) no.7, 1393-1411
doi:10.5506/APhysPolB.45.1393
[arXiv:1406.3658 [hep-ph]].

%\cite{Jegerlehner:2021ckr}
\bibitem{Jegerlehner:2021ckr}
F.~Jegerlehner,
``The Standard Model of particle physics as a conspiracy theory and the possible role of the Higgs boson in the evolution of the early universe,''
Acta Phys. Polon. B \textbf{52} (2021) no.6-7, 575
doi:10.5506/APhysPolB.52.575
[arXiv:2106.00862 [hep-ph]].

%\cite{Wetterich:2011aa}
\bibitem{Wetterich:2011aa}
C.~Wetterich,
``Where to look for solving the gauge hierarchy problem?,''
Phys. Lett. B \textbf{718} (2012), 573-576
doi:10.1016/j.physletb.2012.11.020
[arXiv:1112.2910 [hep-ph]].

%\cite{Dyson:1949ha}
\bibitem{Dyson:1949ha}
F.~J.~Dyson,
``The S matrix in quantum electrodynamics,''
Phys. Rev. \textbf{75} (1949), 1736-1755
doi:10.1103/PhysRev.75.1736.

\bibitem{BS55}
N.~N.~Bogoliubov and D.~V.~Shirkov, ``Problems in quantum field theory. I and II,'' Uspekhi Fiz. Nauk 55, (1955), 149–214 and 57, (1955), 3-92-in Russian; for German translation see Fortschr. d. Physik 3 (1955) 439-495 and 4 (1956) 438-517.

%\cite{Zimmermann:1969jj}
\bibitem{Zimmermann:1969jj}
W.~Zimmermann,
``Convergence of Bogolyubov's method of renormalization in momentum space,''
Commun. Math. Phys. \textbf{15}, 208-234 (1969)
doi:10.1007/BF01645676;

W.~Zimmermann, in 
S.~Deser, M.~Grisaru, H.~Pendleton, Lectures on Elementary Particles and Quantum Field Theory. Volume 1. 1970 Brandeis University Summer Institute in Theoretical Physics, 1970.

%\cite{Martin:1997ns}
\bibitem{Martin:1997ns}
S.~P.~Martin,
``A Supersymmetry primer,''
Adv. Ser. Direct. High Energy Phys. \textbf{18} (1998), 1-98
doi:10.1142/9789812839657\_0001
[arXiv:hep-ph/9709356 [hep-ph]].


%\cite{Giudice:2013yca}
\bibitem{Giudice:2013yca}
G.~F.~Giudice,
``Naturalness after LHC8,''
PoS \textbf{EPS-HEP2013} (2013), 163
doi:10.22323/1.180.0163
[arXiv:1307.7879 [hep-ph]].

%\cite{Cohen:2019wxr}
\bibitem{Cohen:2019wxr}
T.~Cohen,
``As Scales Become Separated: Lectures on Effective Field Theory,''
PoS \textbf{TASI2018} (2019), 011
[arXiv:1903.03622 [hep-ph]].


%\cite{Craig:2022eqo}
\bibitem{Craig:2022eqo}
N.~Craig,
``Naturalness: past, present, and future,''
Eur. Phys. J. C \textbf{83} (2023) no.9, 825
doi:10.1140/epjc/s10052-023-11928-7
[arXiv:2205.05708 [hep-ph]].

%\cite{Appelquist:1974tg}
\bibitem{Appelquist:1974tg}
T.~Appelquist and J.~Carazzone,
``Infrared Singularities and Massive Fields,''
Phys. Rev. D \textbf{11}, 2856 (1975)
doi:10.1103/PhysRevD.11.2856;

%\cite{Symanzik:1973vg}
%\bibitem{Symanzik:1973vg}
K.~Symanzik,
``Infrared singularities and small distance behavior analysis,''
Commun. Math. Phys. \textbf{34} (1973), 7-36
doi:10.1007/BF01646540

%\cite{Schimert:1983fb}
\bibitem{Schimert:1983fb}
T.~Schimert and C.~C.~Chiang,
``HEAVY MASS POWER COUNTING THEOREM AND A PROOF OF DECOUPLING IN THE MOMENTUM SUBTRACTION SCHEME,''
Phys. Rev. D \textbf{29} (1984), 241
doi:10.1103/PhysRevD.29.241

%\cite{Caswell:1981ek}
\bibitem{Caswell:1981ek}
W.~E.~Caswell and A.~D.~Kennedy,
``A SIMPLE APPROACH TO RENORMALIZATION THEORY,''
Phys. Rev. D \textbf{25} (1982), 392
doi:10.1103/PhysRevD.25.392

%\cite{Kazama:1981fx}
\bibitem{Kazama:1981fx}
Y.~Kazama and Y.~P.~Yao,
``Decoupling, Effective Lagrangian, and Gauge Hierarchy in Spontaneously Broken Nonabelian Gauge Theories,''
Phys. Rev. D \textbf{25} (1982), 1605
doi:10.1103/PhysRevD.25.1605



%\cite{Nambu:1968rr}
\bibitem{Nambu:1968rr}
Y.~Nambu,
``S Matrix in semiclassical approximation,''
Phys. Lett. B \textbf{26} (1968), 626-629
doi:10.1016/0370-2693(68)90436-X

%\cite{Peskin:1995ev}
\bibitem{Peskin:1995ev}
M.~E.~Peskin and D.~V.~Schroeder,
``An Introduction to quantum field theory,''
Addison-Wesley, 1995,
ISBN 978-0-201-50397-5

%\cite{Weinberg:1995mt}
\bibitem{Weinberg:1995mt}
S.~Weinberg,
``The Quantum theory of fields. Vol. 1: Foundations,''
Cambridge University Press, 2005,
ISBN 978-0-521-67053-1, 978-0-511-25204-4
doi:10.1017/CBO9781139644167

%\cite{Wilson:1971bg}
\bibitem{Wilson:1971bg}
K.~G.~Wilson,
``Renormalization group and critical phenomena. 1. Renormalization group and the Kadanoff scaling picture,''
Phys. Rev. B \textbf{4}, 3174-3183 (1971)
doi:10.1103/PhysRevB.4.3174;

%\cite{Wilson:1971dh}
%\bibitem{Wilson:1971dh}
K.~G.~Wilson,
``Renormalization group and critical phenomena. 2. Phase space cell analysis of critical behavior,''
Phys. Rev. B \textbf{4}, 3184-3205 (1971)
doi:10.1103/PhysRevB.4.3184

%\cite{Coleman:2018mew}
\bibitem{Coleman:2018mew}
S.~Coleman, B.~G.~g.~Chen, D.~Derbes, D.~Griffiths, B.~Hill, R.~Sohn and Y.~S.~Ting,
``Lectures of Sidney Coleman on Quantum Field Theory,''
WSP, 2018,
ISBN 978-981-4632-53-9, 978-981-4635-50-9
doi:10.1142/9371

%\cite{Georgi:1993mps}
\bibitem{Georgi:1993mps}
H.~Georgi,
``Effective field theory,''
Ann. Rev. Nucl. Part. Sci. \textbf{43} (1993), 209-252
doi:10.1146/annurev.ns.43.120193.001233

%\cite{Choi:2025tus}
\bibitem{Choi:2025tus}
K.~S.~Choi,
``Self-Similar Structure of Loop Amplitudes and Renormalization,''
[arXiv:2502.19300 [hep-th]];

%\cite{Choi:2025mho}
%\bibitem{Choi:2025mho}
K.~S.~Choi,
``Self-Similarity of Loop Amplitudes,''
[arXiv:2503.14330 [hep-th]].

%\cite{Collins:1984xc}
\bibitem{Collins:1984xc}
J.~C.~Collins,
``Renormalization : An Introduction to Renormalization, the Renormalization Group and the Operator-Product Expansion,''
Cambridge University Press, 1984,
ISBN 978-0-521-31177-9, 978-0-511-86739-2, 978-1-009-40180-7, 978-1-009-40176-0, 978-1-009-40179-1
doi:10.1017/9781009401807

\bibitem{Bogoliubov:1957gp}
N.~Bogoliubov and O.~Parasiuk, 
\textit{{
\"Uber die Multiplikation der Kausalfunktionen in der Quantentheorie der Felder}},
Acta Math. \textbf{97}
  (1957) 227--266.

\bibitem{Hepp:1966eg}
K.~Hepp,
 \textit{{Proof of the Bogolyubov-Parasiuk theorem on renormalization}}, 
Commun. Math. Phys. \textbf{2} (1966) 301--326.

\bibitem{Zimmermann:1967}
W. Zimmermann,
\textit{Local field equation for $A^{4}$-coupling in renormalized perturbation theory},
Commun. Math. Phys. \textbf{6} (1967) 161--188.

%\cite{Lowenstein:1975ps}
\bibitem{Lowenstein:1975ps}
J.~H.~Lowenstein,
``Convergence Theorems for Renormalized Feynman Integrals with Zero-Mass Propagators,''
Commun. Math. Phys. \textbf{47} (1976), 53-68
doi:10.1007/BF01609353


%\cite{Lowenstein:1975rg}
\bibitem{Lowenstein:1975rg}
J.~H.~Lowenstein and W.~Zimmermann,
``The Power Counting Theorem for Feynman Integrals with Massless Propagators,''
Commun. Math. Phys. \textbf{44} (1975), 73-86
doi:10.1007/BF01609059


%\cite{Weinberg:1959nj}
\bibitem{Weinberg:1959nj}
S.~Weinberg,
``High-energy behavior in quantum field theory,''
Phys. Rev. \textbf{118}, 838-849 (1960)
doi:10.1103/PhysRev.118.838.

\bibitem{BS}
N.~N.~Bogoliubov and D.~V.~Shirkov,  ``Introduction to the Theory of Quantized Fields.''   John Wiley \& Sons Inc; 3rd edition (1980).

%\cite{Blaschke:2013cba}
\bibitem{Blaschke:2013cba}
D.~N.~Blaschke, F.~Gieres, F.~Heindl, M.~Schweda and M.~Wohlgenannt,
``BPHZ renormalization and its application to non-commutative field theory,''
Eur. Phys. J. C \textbf{73} (2013), 2566
doi:10.1140/epjc/s10052-013-2566-8
[arXiv:1307.4650 [hep-th]].


%\cite{Georgi:1976vf}
\bibitem{Georgi:1976vf}
H.~Georgi and H.~D.~Politzer,
``Precocious Scaling, Rescaling and xi Scaling,''
Phys. Rev. Lett. \textbf{36} (1976), 1281
[erratum: Phys. Rev. Lett. \textbf{37} (1976), 68]
doi:10.1103/PhysRevLett.36.1281

%\cite{Kraus:1997bi}
\bibitem{Kraus:1997bi}
E.~Kraus,
``Renormalization of the Electroweak Standard Model to All Orders,''
Annals Phys. \textbf{262} (1998), 155-259
doi:10.1006/aphy.1997.5746
[arXiv:hep-th/9709154 [hep-th]].

%\cite{tHooft:1979rat}
\bibitem{tHooft:1979rat}
G.~'t Hooft,
``Naturalness, chiral symmetry, and spontaneous chiral symmetry breaking,''
NATO Sci. Ser. B \textbf{59} (1980), 135-157
doi:10.1007/978-1-4684-7571-5\_9


%\cite{tHooft:1972tcz}
\bibitem{tHooft:1972tcz}
G.~'t Hooft and M.~J.~G.~Veltman,
``Regularization and Renormalization of Gauge Fields,''
Nucl. Phys. B \textbf{44}, 189-213 (1972)
doi:10.1016/0550-3213(72)90279-9.

%\cite{Bollini:1972ui}
\bibitem{Bollini:1972ui}
C.~G.~Bollini and J.~J.~Giambiagi,
``Dimensional Renormalization: The Number of Dimensions as a Regularizing Parameter,''
Nuovo Cim. B \textbf{12}, 20-26 (1972)
doi:10.1007/BF02895558


%\cite{Dirac:1937ti}
\bibitem{Dirac:1937ti}
P.~A.~M.~Dirac,
``The Cosmological constants,''
Nature \textbf{139} (1937), 323
doi:10.1038/139323a0

%\cite{Yamada:2020cvt}
\bibitem{Yamada:2020cvt}
M.~Yamada,
``Gauge hierarchy problem and scalegenesis,''
PoS \textbf{CORFU2019} (2020), 077
doi:10.22323/1.376.0077
[arXiv:2004.00142 [hep-ph]].

%\cite{Branchina:2022gll}
\bibitem{Branchina:2022gll}
C.~Branchina, V.~Branchina and F.~Contino,
``Physical tuning and naturalness,''
Phys. Rev. D \textbf{107} (2023) no.9, 096012
doi:10.1103/PhysRevD.107.096012
[arXiv:2208.05431 [hep-ph]].


%\cite{Brivio:2017vri}
\bibitem{Brivio:2017vri}
I.~Brivio and M.~Trott,
``The Standard Model as an Effective Field Theory,''
Phys. Rept. \textbf{793} (2019), 1-98
doi:10.1016/j.physrep.2018.11.002
[arXiv:1706.08945 [hep-ph]].

%\cite{Bardeen:1995kv}
\bibitem{Bardeen:1995kv}
W.~A.~Bardeen,
``On naturalness in the standard model,''
FERMILAB-CONF-95-391-T.



%\cite{Sirlin:1980nh}
\bibitem{Sirlin:1980nh}
A.~Sirlin,
``Radiative Corrections in the $\mathrm{SU}(2)_L \times \mathrm{U}(1)$ Theory: A Simple Renormalization Framework,''
Phys. Rev. D \textbf{22} (1980), 971-981
doi:10.1103/PhysRevD.22.971

%\cite{Ambjorn:1979nv}
\bibitem{Ambjorn:1979nv}
J.~Ambj{\o}rn,
``On the decoupling of massive particles in field theory,''
Commun. Math. Phys. \textbf{67} (1979), 109-119
doi:10.1007/BF01221360

%\cite{Manoukian:1981pb}
\bibitem{Manoukian:1981pb}
E.~B.~Manoukian,
``Decoupling theorem in field theory,''
J. Math. Phys. \textbf{22} (1981), 572;

E.~B.~Manoukian,
``Generalized decoupling theorem in quantum field theory,''
J. Math. Phys. \textbf{22} (1981), 2258-2262
doi:10.1063/1.524797;

E.~B.~Manoukian,
``Proof of the decoupling theorem of field theory in Minkowski space,''
J. Math. Phys. \textbf{25} (1984), 1519-1523
doi:10.1063/1.526324

%\cite{Manoukian:1983book}
\bibitem{Manoukian:1983book}
E.~B.~Manoukian,
``Renormalization,''
Academic Press, New York, 1983,
ISBN 0-12-469450-0

%\cite{Gorishnii:1989dd}
\bibitem{Gorishnii:1989dd}
S.~G.~Gorishnii,
``Construction of operator expansions and effective theories in the MS scheme,''
Nucl. Phys. B \textbf{319} (1989), 633-666

%\cite{Smirnov:1990rz}
\bibitem{Smirnov:1990rz}
V.~A.~Smirnov,
``Asymptotic expansions in limits of large momenta and masses,''
Commun. Math. Phys. \textbf{134} (1990), 109-137

%\cite{Smirnov:2002pj}
\bibitem{Smirnov:2002pj}
V.~A.~Smirnov,
``Applied Asymptotic Expansions in Momenta and Masses,''
Springer Tracts Mod. Phys. \textbf{177} (2002), 1-262


\end{thebibliography}
\end{document}